# Artificial Intelligence, Lean Startup Method, and Product Innovations

Xiaoning Wang

Jindal School of Management,
University of Texas at Dallas

Lynn Wu

The Wharton School,
University of Pennsylvania

**Abstract:** Although artificial intelligence (AI) has the potential to drive significant business innovation, many firms struggle to realize its benefits. We investigate why some firms succeed in using AI for innovation while others fail, focusing on the organizational support necessary for leveraging AI in both novel and incremental innovation. Specifically, we examine how the Lean Startup Method (LSM) influences the impact of AI on product innovation in startups. Analyzing data from 1,800 Chinese startups between 2011 and 2020, alongside policy shifts by the Chinese government in encouraging AI adoption, we find that companies with strong AI capabilities produce more innovative products. Moreover, our study reveals that AI investments complement LSM in innovation, with effectiveness varying by the type of innovation and AI capability. We differentiate between discovery-oriented AI, which reduces uncertainty in novel areas of innovation, and optimization-oriented AI, which refines and optimizes existing processes. Within the framework of LSM, we further distinguish between prototyping—focused on developing minimum viable products—and controlled experimentation—focused on rigorous testing such as A/B testing. We find that LSM complements discovery-oriented AI by utilizing AI to expand the search for market opportunities and employing prototyping to validate these opportunities, thereby reducing uncertainties and facilitating the development of the first release of products. Conversely, LSM complements optimization-oriented AI by using A/B testing to experiment with the universe of input features and using AI to streamline iterative refinement processes, thereby accelerating the improvement of iterative releases of products. As a result, when firms use AI and LSM for product development, they are able to generate more high-quality products in less time. These findings, applicable to both software and hardware development, underscore the importance of treating AI as a heterogeneous construct because different AI capabilities require distinct organizational processes to achieve optimal outcomes.

**Keywords:** Artificial Intelligence, Lean Startup Method, Startup, Product Development, Innovation.

## 1. Introduction and Motivation

New product development involves activities that begin with identifying a market opportunity and culminate in creating a product or service that meets market needs (Krishnan and Loch 2005). These activities are crucial for startups to capture the potential market gains of successful products. However, innovation is risky, as failed products can be costly and even lead to bankruptcy (Bhaskaran et al. 2021). One of the most well-known methodologies for developing new products is the lean startup method (LSM)[1], which is used widely in both early-stage and mature firms (Koning et al. 2022, Yoo et al. 2021). Inspired by the concept of "lean manufacturing" from the Toyota Production System, LSM emphasizes validating new products through market feedback and continuous experimentation, rather than relying solely on business plans or entrepreneurial insights (Ries 2011, Blank 2013). For example, startups can build prototypes to gather consumer feedback and iteratively improve their products, pivoting if necessary.

Despite its advantages, LSM can increase marketing and operating costs, which may deter some startups from adopting it (Felin et al. 2020, Bhamu and Sangwan 2014, Anand et al. 2016). As AI technologies continue to advance, many startups are incorporating them into product development (Berente et al. 2021, Verganti et al. 2020). Some firms use AI to analyze market data and uncover new insights. For example, Genki Forest, a Chinese startup initially focused on tea beverages, used AI to analyze consumer data on major social platforms. They discovered that young people in China were highly concerned about sugar, diabetes, and obesity in their drink choices, leading Genki Forest to create zero-sugar sparkling water. This product was highly successful because AI enabled the company to identify a new market and launch a product tailored to it.[2] Others, like Airbnb, use AI to automate existing business processes, such as converting hand-drawn sketches into source code, to accelerate product iterations.[3]

[1] Despite having the word "startup" in the middle of its name, these principles of LSM are widely embraced by both startups and mature firms. The use of this term does not imply our findings only apply to startups. Future research may further explore the application of our findings to companies past the startup phase.

However, research shows that most companies have not fully benefited from AI despite significant investments. Simply adopting AI is not enough; its successful use requires tailored organizational support aligned with the specific type of AI being implemented (Varian 2018, Brynjolfsson et al. 2021, Dixon et al. 2021). In this paper, we identify LSM as a crucial form of organizational support and explore how AI can either substitute or complement LSM in product innovation. Understanding this relationship can help explain why some startups successfully use AI to innovate while others do not.

Using a panel dataset of 1,800 startups in China from 2011 to 2020, we find that startups with AI capabilities can develop more products, including both first releases (defined as new products for the company) and iterative releases (defined as new versions of existing products). Furthermore, LSM is complementary to AI capabilities in developing more products in less time, and the benefit adopting one without the other is limited.

To explore the underlying mechanisms, we differentiate between discovery-oriented AI, which discovers new insights, and optimization-oriented AI, which optimizes existing processes. We borrow the well-documented concepts of IT ambidexterity, where exploration-oriented IT refers to employing IT resources to explore new practices, and exploitation-oriented IT refers to refining known IT practices (March 1991, Levinthal and March 1993, Lee at al. 2015, Liang et al. 2022, Gregory et al. 2015). Building on this literature, we characterize discovery-oriented AI as the ability to generate new insights, identify patterns, and form hypotheses from data, with the goal of reducing uncertainty about the unknown. Examples include using machine learning models for market predictions and demand forecasting. Optimization-oriented AI, on the other hand, focuses on optimizing, automating, and refining existing processes, products, and strategies based on known insights and patterns. Examples include robotic process automation (RPA) and AI-driven workflow optimization.

[2] For more information, see https://techcrunch.com/2021/07/25/data-driven-iteration-helped-chinas-genki-forest-become-a-6b-beverage-giant-in-5-years/.
[3] See https://airbnb.design/sketching-interfaces/.

To examine how different types of AI complement LSM, we also distinguish between two key practices within LSM: prototyping (e.g., developing minimum viable products) and controlled experimentation (e.g., A/B testing). Our findings indicate that (1) prototyping complements discovery-oriented AI in developing first releases but not iterative ones, and (2) controlled experimentation complements optimization-oriented AI in refining iterative releases but not first ones. These results suggest that although AI and LSM are complementary in product innovations, how specific aspects of LSM and AI are paired is critical to unlock the potential of AI in product innovation. The findings remain robust even when using local government policies as an exogenous shock for AI adoption and employing alternative measurements of key constructs.

In summary, our results highlight the importance of aligning AI capabilities with appropriate organizational supports like LSM. Differentiating AI types is crucial for understanding their impact on product innovation, and the correct matching of AI capabilities with the necessary organizational complements is key to fostering product development and innovation.

## 2. Literature and Theory

In this section, we first review the literature about product development and how startups use LSM for product innovations. Then we theorize the complementarities between AI capabilities and LSM as well as the complementarities between specific components of each.

### 2.1 Startup Product Innovation and LSM

Recent decades witnessed the remarkable impact of startups on product innovation that disrupted many industries (Tambe 2014, Liu et al. 2016, Ransbotham et al. 2019, Hellmann and Thiele 2015, Burtch et al. 2016, Mei et al. 2022 Gompers and Lerner 2001, Brynjolfsson and McAfee 2014). Even the disrupters are being disrupted as fierce competition rages among startups in part due to rapid changes in the technology landscape. To adapt to the fast-changing technology and market dynamics, many startups and others adopt LSM for product innovation (Blank 2005, Ries 2011, Blank 2013, Rigby et al. 2018).

Unlike rigid adherence to a predefined business plan or entrepreneurial intuition, LSM emphasizes validating ideas in the marketplace and iteratively improving products based on feedback. This approach has gained widespread adoption across software and hardware industries (Sarasvathy 2001, Read et al. 2011).

The literature has identified two primary applications of LSM: prototyping and controlled experimentation (Yoo et al. 2021, Harms and Schwery 2020, Calantone et al. 2002, Koning et al. 2022, Cui and Wu 2017). Prototyping focuses on developing minimum viable products (MVPs) to gather consumer insights and address uncertainties before a product launch (Yoo et al. 2021, Camuffo et al. 2020). Controlled experimentation, in contrast, involves systematically testing and refining specific features of established products to optimize performance (Kohavi et al. 2020, Koning et al. 2022, Thomke 2020).

These approaches differ in timing, objectives, and required expertise. Prototyping typically involves qualitative methods, such as interviews or surveys, to collect contextual market insights, necessitating industry-specific expertise. However, contextual data is inherently difficult to transfer across settings and challenging to integrate with broader market data, requiring careful analysis and contextural insights to prevent misinterpretation (von Hippel 1994).

Controlled experimentation, by comparison, leverages digital traces generated by existing products to process large datasets using statistical methods. This requires expertise in experimental design and data analysis, which are broadly applicable across industries. While prototyping focuses on reducing market uncertainty and validating product-market fit, controlled experimentation aims to refine and optimize product performance through data-driven approaches.

Despite its benefits, LSM is not without challenges. Prototyping can be time-intensive, requiring substantial effort to narrow down to the optimal market niches, while controlled experimentation demands significant operational resources. Moreover, LSM's focus on responding to immediate market feedback can lead to prioritizing incremental innovations over long-term, radical breakthroughs (Felin et al. 2020, Thiel and Masters 2014, Mollick 2019, Bhamu and Sangwan 2014, Felin and Foss 2011, Anand et al. 2016).

To address these limitations, firms increasingly adopt advanced technologies like AI to reduce costs and enhance product innovation. AI tools enable faster, more accurate analyses of market and user data, helping startups strike a balance between short-term responsiveness and long-term strategic goals.

**2.2 AI and Product Innovation**

In this study, we align with recent literature defining AI as advanced computing power for making predictive actions using machine learning techniques, often emphasizing high accuracy at the expense of interpretability (Agrawal et al. 2018, Taddy 2018, Berente et al. 2021, Brynjolfsson et al. 2021). The "black-box" nature of AI algorithms highlights the importance of understanding the context of the data being used (Berente et al. 2021, Li et al. 2021, Taddy 2018). As with prior generations of information technology, AI is multifaceted and multi-purposed (Bresnahan and Trajtenberg 1995, Varian 2018, Berente et al. 2021, Brynjolfsson et al. 2021). To better understand how different aspects of AI influence product innovation in startups, we categorize general AI capabilities into two types: discovery-oriented AI and optimization-oriented AI.

**Discovery-oriented AI** focuses on uncovering new areas of innovation by analyzing large datasets to reveal hidden patterns, trends, and generate hypotheses. This type of AI is instrumental in identifying product directions, validating market demand, and increasing certainty about the correctness of the objective function. Professionals leveraging discovery-oriented AI, such as machine learning analysts, typically focus on conducting market research and exploring new product development opportunities.

In contrast, **optimization-oriented AI** is designed to refine, automate, and optimize existing processes, products, and strategies where the objective function is already established. This type of AI leverages advanced techniques to enhance efficiency and performance. Professionals in this domain often include RPA developers, AI operations engineers, and workflow automation specialists who focus on improving existing processes.

The two types of AI differ in their objectives and the contextual knowledge required to operate effectively. Discovery-oriented AI addresses uncertainty in undefined, exploratory domains, where goals are often ambiguous and require deep domain expertise. Practitioners using this type of AI must have a thorough understanding of industry-specific challenges to frame relevant questions, identify key variables, design exploratory models, and interpret results accurately (Lebovitz et al. 2021).

In contrast, optimization-oriented AI is more focused on enhancing efficiency in well-defined and structured tasks. Unlike discovery-oriented AI, this approach is less industry-specific and is adaptable across various sectors. By leveraging structured datasets, optimization-oriented AI is able to automate processes and maximize outputs with limited customization. For instance, robotic process automation (RPA) often employs AI to handle routine tasks such as data entry, invoice processing, or customer service. These tasks are typically standardized across industries, enabling optimization-oriented AI to deliver results with minimal adjustments.

In the context of product innovation, discovery-oriented AI and optimization-oriented AI address different challenges. Developing new products involves overcoming the uncertainty of customer demand, as predicting whether customers will adopt a new product remains inherently challenging (Lynn et al. 1996, Hoeffler 2003, Kwark et al. 2018). Discovery-oriented AI plays a critical role in this process by rapidly analyzing vast amounts of market information, predicting market dynamics, and identifying new business opportunities. This capability expands the scope of potential market opportunities while reducing uncertainties. For example, Stitch Fix, a fashion retail company that offers online personal styling services, utilizes AI algorithms to predict granular shifts in apparel demand by analyzing millions of customer style profiles, purchase histories, feedback ratings, and emerging trends from external data sources, including social media and fashion blogs. This enables the company to forecast which clothing categories, colors, and silhouettes will resonate with specific demographic and geographic segments months before each season's production cycle, thereby reducing uncertainty in new collection launches. By processing this high-dimensional data, Stitch Fix's AI system can generate and evaluate a far larger variety of style combinations than human stylists alone, broadening the range of apparel designs the company brings to market and improving the likelihood of first-release success.

When refining an existing product, the focus shifts from gauging uncertain demand to iterating quickly to improve quality. This process involves fine-tuning product features and optimizing existing development processes, rather than exploring unknown markets (Wilson and Norton 1989, Netessine and Taylor 2007, Hashai and Markovich 2017, Gans et al. 2002).

Optimization-oriented AI is well-suited for this context, as it automates repetitive tasks and enhances iteration efficiency. For instance, Turing Labs, a startup, employs AI to automate product formulation processes, significantly reducing the workload of technicians and improving iteration speed. According to the company, “the process was 90% manual but now is 90% digital.” By streamlining development processes and removing scaling limitations, optimization-oriented AI enables faster product iterations and higher efficiency in refining existing offerings.

Given that discovery-oriented AI supports the creation of first-release products and optimization-oriented AI enhances the improvement of iterative releases, we hypothesize that AI use can facilitate product innovation in startups by helping startups shorten the product development cycle as well increase the number of products they can develop.

*H1: Using AI can raise the quantity of a startup’s product innovations.*

**2.3 Complementarity**

Despite the potential benefits of AI in fostering innovation and productivity, many organizations that invested in AI failed to realize its benefits (Lou and Wu 2021, Berente et al. 2021). A key factor contributing to this challenge is the "black-box" nature of AI, which limits its utility when dealing with sparse or unreliable data (Lou and Wu 2021, Wu et al. 2019). This issue becomes particularly problematic in the context of innovation, where reliable data is often scarce. Recent literature has examined how biases in input data can lead to “poor learning” (Cao et al. 2022, Hoelzemann et al. 2022), potentially exacerbating information distortion and negatively impacting a company’s innovation performance.[4] Paradoxically, the more innovative an idea, the less likely reliable data is available (Thomke 2020). For example, using AI to predict market demand for a completely new product may be ineffective when no historical data exists, and consumers struggle to conceptualize how the product could be used (Cooper et al. 2002). Even in iterative product development, where market uncertainty is lower, technological uncertainty persists. Defining "higher quality" can remain ambiguous, and available data is often limited (Allon et al. 2021, Krishnan and Bhattacharya 2002, Lauga and Ofek 2009). This complexity requires frequent calibration and improvement of AI algorithms, supported by complementary organizational processes to meet long-term product innovation needs.

[4] We are grateful to anonymous reviewers for bringing this up.

LSM can serve as a crucial organizational complement to AI in product innovation. While both AI and LSM are widely used, their approaches to navigating market challenges differ significantly. AI relies on analyzing existing data to generate ideas, derive insights and predict outcomes. In contrast, LSM focuses on generating new data through prototyping and experimentation, relying on customer feedback to validate hypotheses and refine products. This difference in approach is particularly useful in situations where data is sparse or unreliable. In such cases, LSM can produce firsthand, real-world data to validate and refine AI-generated insights, improving the performance of AI algorithms over time and mitigating the risk of "poor learning". Conversely, AI can accelerate the LSM process by identifying new opportunities and automating repetitive tasks (Kerr et al. 2014; Ewens et al. 2018; Camuffo et al. 2020).

When combined, AI and the Lean Startup Methodology (LSM) offer a powerful, complementary framework for product innovation. Generative AI, pretrained on vast internet-scale data, can dramatically expand the search space, generating a greater number and diversity of ideas than humans alone (Meinche et al., 2025). However, selecting the most promising ideas and refining them requires real-time market feedback. This is where LSM plays a crucial role—providing real-world validation and fresh data that AI can further analyze. By leveraging AI's analytical capabilities alongside LSM's iterative approach, startups can quickly identify product-market fit, optimize design features, and minimize the trial-and-error inherent in traditional development. LSM's focus on building minimum viable products (MVPs) and testing them through continuous experimentation aligns naturally with AI's strength in processing large datasets, predicting trends and generating product ideas. Together, these approaches broaden the ideation space, accelerate hypothesis testing, improve decision-making with data-driven insights, and reduce the time and resources needed to develop and launch successful products. This complementariy enables startups not only to build more products but also to bring them to market faster. Therefore, we hypothesize:

*H2: AI capability complements LSM for product innovations.*

*2.3.1 Discovery-oriented AI complements prototyping for generating first releases*

The complementarity between AI and LSM becomes even more evident when considering specific aspects of both methodologies, particularly the roles of prototyping and controlled experimentation in LSM, as well as discovery-oriented and optimization-oriented AI. These distinctions are crucial for understanding how AI and LSM align in different product development stages. Aligning R&D objectives and required contextual knowledge and skills in different stages is essential for identifying and explaining their complementarities.

Developing the first release is primarily in the early stage, where the major challenge is to find the right market niche. Prototyping, an integral part of early-stage development, involves testing hypotheses to understand customer needs and reduce market uncertainty. Similarly, discovery-oriented AI is also deployed at this stage to analyze unknown market dynamics, consumer behavior, and trends. Both approaches share the objective of reducing market uncertainties and strategically positioning a product. They are key to developing first releases.

Discovery-oriented AI requires domain-specific knowledge to frame relevant questions,

design exploratory models, and interpret results effectively. This aligns with the industry-specific expertise needed for prototyping. Together, they work to validate hypotheses and refine product ideas. In developing first releases, discovery-oriented AI broadens the search for potential business opportunities, guides where to launch prototypes, and focuses efforts on high-potential markets.

Because discovery-oriented AI and prototyping are aligned in the product development objectives and required contextual knowledge,  discovery-oriented AI can significantly accelerate the prototyping process. Generative AI, in particular, simplifies the ideation phase using pre-trained transformers that allow teams to quickly generate and test a variety of concepts. Launching prototypes in the wild using LSM can further acquire the data that discovery-oriented AI needs to further improve or pivot the prototype. For instance, advertising professionals can use AI-driven image generation tools to create diverse personalized ads that can be tested in real-world scenarios (Chen and Chan 2024). Once prototypes of these ads are launced for testing and consumer behaviors are observed, discovery-oriented AI can generate valuable insights about market dynamics and preferences. These insights help narrow down prototype selections, making the entire process more targeted and efficient. By integrating discovery-oriented AI with prototyping, organizations can effectively reduce market uncertainty and increase the likelihood of identifying and successfully launching new products.

The complementarity between AI and prototyping extends to both digital and physical products. For example, Genki Forest used AI to launch several low-sugar test products. Initially focused on zero-sugar tea beverages, market feedback from prototypes led to a strategic pivot, resulting in the best-selling zero-sugar sparkling water. In dynamic markets, discovery-oriented AI expands the scope for creating prototypes, enabling the development of products that traditional methods might miss.

*H3a: There are positive interaction effects between discovery-oriented AI capability and prototyping in developing first releases in startups.*

*2.3.2 Optimization-oriented AI complements controlled experimentation for generating iterative releases*

Iterative release development typically occurs in the later stages of product development, where the main objective shifts to refining and optimizing an existing product. The primary goal at this stage is to address known customer needs and enhance product features to maintain competitiveness in a relatively stable market. Controlled experimentation and optimization-oriented AI are aligned because both share the same goal of expediting product iterations and enhancements. Controlled experimentation is used to iteratively improve existing products once the market is validated. Similarly, optimization-oriented AI focuses on automating existing processes and enhancing operational efficiency.

Because of the shared goal of automating existing processes and enhancing operational efficiency based on gathered data, optimization-oriented AI can complement controlled experimentation. Optimization-oriented AI automates routine tasks and optimizes established processes, significantly speeding up product iterations. Controlled experimentation complements optimization-oriented AI by testing different input features in product development, providing feedback data to further refine AI algorithms. These data can then be fed back to optimization-oriented AI to further refine feature selection. For example, Airbnb employs optimization-oriented AI to automate the classification of room images, streamlining the process for hosts and enhancing the user experience. This capability enabled Airbnb to launch Airbnb Plus quickly. To improve the AI algorithm's efficiency, Airbnb used A/B testing, a type of controlled experimentation, to identify the most relevant image features, which simplified the AI's processing requirements and enhanced its performance. Controlled experimentation provided critical feedback that refined the AI algorithm, demonstrating the complementary relationship between these methodologies.

Furthermore, AI can optimize the experimentation process itself. Advanced AI techniques, such as multi-armed bandit algorithms, are often used to maximize the value of experimental data, particularly when sample sizes are limited. By combining such techniques with controlled experimentation, firms can significantly enhance the efficiency and effectiveness of product iterations.

In summary, discovery-oriented AI complements prototyping by addressing market uncertainties and facilitating the creation of first releases. Optimization-oriented AI complements controlled experimentation by improving iteration efficiency and refining existing products. Therefore, we hypothesize:

*H3b: There are positive interaction effects between optimization-oriented AI capability and controlled experimentation in developing iterative releases in startups.*

However, while we anticipate that discovery-oriented AI will complement prototyping and optimization-oriented AI will complement controlled experimentation, we do not expect strong complementarities between discovery-oriented AI and controlled experimentation or between optimization-oriented AI and prototyping. This is because these cross-complementarities do not align in terms of R&D stages, shared objectives, and contextual knowledge. Proper alignment ensures that the appropriate tools and methodologies are applied to achieve intended purposes.

## 3. Data and Measures

### 3.1 Company Information Data

We study startups as they are at the frontiers of product innovation. To compete with incumbents, startups are incentivized to adopt nascent technologies such as AI in their innovation activities. We acquired startup information from Crunchbase, one of the most widely used databases for global entrepreneurial research (Cumming et al. 2016, Wang et al. 2023). We examine startups that were founded between 2011 and 2020, the same time frame as many recent studies that examine firm-level AI adoption (Lou and Wu 2021, Babina et al. 2024). We focus on companies in China because China has been a leading economy in AI adoption, especially since 2014, when the Chinese government started to issue AI-related policies to encourage AI adoption in firms. In total, there are about 1,800 Chinese startups, including about 600 software companies and 1,200 companies that develop physical products.[5] For each of these startups, we obtain the founding year and specific funding outcomes—the number of successful rounds, the amount of financing received, and the financing sources in all funding series (Seed, Series A, Series B, etc.). The average number of funding rounds is about 1.5. We match these startups with the Chinese Enterprise Database. The database curates comprehensive information about all the registered companies in China, from public data sources and proprietary governmental databases, including location, industry category, subsidiary and branch information, job postings on major hiring websites, yearly personnel information, WeChat public accounts, founder and board member information, the detailed administrative information from all governmental departments, the detailed certificate, trademark, patent and copyright information, and also all the online news reports that relate to the focal company from over 100 major news agencies in China. This data is validated through other similar data sources such as *Tianyancha*.

Using the company's WeChat public accounts, we acquired all the historical articles published in each account, including the title, text, and the time of publication for each article, and whether the article is an original article published by the company or a reposted article. WeChat is the most widely used social media platform in China and nearly all firms have a presence on it. Firms use WeChat heavily to make public announcements and broadcast product information, which can help us infer the company's adoption level of AI capability and LSM.

[5] Data acquired in 2020.

### 3.2 Variables

***Dependent Variables***

For software companies, we measure the number of new products using the **Number of Software Copyrights** that a company applies for each year. For physical product companies, we use the **Number of Product Trademarks** that a company applies for each year (Babina et al. 2024, Gao and Hitt 2012). Software copyrights and product trademarks measure product development outcomes rather than the development process. These numbers are useful because they are common practices for companies in China to protect intellectual property before launching new products. An example of software copyrights and product trademarks is shown in Figure 1.

<Figure 1 Inserted Here>

To identify whether the software is a first release, we collect the registered software version from the copyright certificate. To register a software copyright, the assignee needs to declare the version number of its software, such as "V1.0" or "V4.3". We treat all "V1.0" software as the **first release** of the firm and subsequent numbers as **iterative release** versions of an existing product. To identify whether a physical product is a first release, we collect the registered product sub-classes from the trademark certificates. To register a trademark, the assignee needs to declare the product sub-classes, and most trademarks have over 10 different sub-classes. We treat a trademark as **a first release** if it contains at least one new sub-class in comparison to the company's existing trademarks, as it suggests the company is expanding to a new market niche, and the rest are treated as **iterative releases**.

We also calculate the **Number of High-Quality Products** for each firm each year to alleviate the possibility that some firms use AI and LSM to generate low-value innovation. This variable measure is based on the number of product certificates issued by independent third-party certification agencies such as the International Organization for Standardization (ISO)[6]. These certifications of quality by neutral third parties are a good indicator of high-quality products.

One concern about the dependent variables (software copyright and product trademark) is that they are not representative of a startup's overall productivity. To measure the overall performance of startups, traditional metrics such as revenue and profit are not publicly available and are not verified by public accountants. To alleviate this concern, we also use the **log Amount of Funding** as the dependent variable to further verify the robustness of the results, as funding is one of the most important financial indicators of startup performance (Bardhan et al. 2013, Wang et al. 2023). We also use the **Search Index** of each startup each year on the major search engine in China, as prior literature has shown that online traffic data can be used to predict and verify firm revenue (Froot et al. 2017, Huang 2018, Zhu 2019). The results are listed in Appendix A1 and are consistent with the main results.

***AI Capabilities***

To measure the extent to which a startup can utilize AI resources in research and development, we measure a startup's **AI Capability** by combining AI-related job posts, original social media posts, and news reports about a firm into a single measure[7] and taking the first principal component. AI job posts have been used in previous studies to measure a company's AI capability (Lou and Wu 2021, Babina et al. 2024). We complement this measure by using social media posts and news articles to measure a firm's overall AI capability. We identify a job post, social media post, and news report as AI-related if it contains any of the AI keywords (Table 1) in the summary texts. We then separately estimate each of the three AI metrics into our main regression and find that we cannot reject the hypothesis that all three measures have the same coefficients. We further validate these text-based measurements by showing that these AI-related job postings, social media posts, and news reports are significantly correlated with firms' actual AI capability measured by AI patents (Lou and Wu 2021). The results are in Appendix A2.[8]

[6] We excluded the China Compulsory Certifications (3C) as an independent certification authority.
[7] Cumulative quantity with a yearly depreciation rate of 15% following literature convention (Lou and Wu 2021). The same is true for the LSM Level below.

<Table 1 Inserted Here>

We also differentiate the startups' discovery-oriented versus optimization-oriented AI capabilities based on the company's usage of AI in discovering new insights and exploring unknowns, or optimizing existing processes and refining knowns. We adopt some widely-used keywords from literature in organizational learning theory, such as "search", "explore", and "discover" for discovery and "refine", "exploit", "efficiency" for optimization (Gregory et al. 2015, March 1991). In addition, we also manually labeled 1,000 randomly selected AI-related job postings, news articles, and social media articles as discovery-oriented or optimization-oriented, and then used a bag-of-word method to identify the discovery and optimization keywords with the greatest coefficients in predicting whether a text is discovery- or optimization-oriented in AI (Table 1). We identify a job post, social media post or news report as discovery AI-related if it contains both AI and discovery keywords within the same sentence in the post, or as optimization AI-related if it contains both AI and optimization keywords.[9] We take the first principal component

[8] We are grateful to the Associate Editor and an anonymous reviewer for bringing this up.

[9] An example of an discovery-oriented AI job posting is: "…conduct explorative market analysis with NLP technique…". The job title is usually Business Intelligence or Market Analyst. An example of an optimization-oriented

to measure a company's **Discovery-oriented AI** and **Optimization-oriented AI**. We also use neural networks to predict whether a text is discovery AI-oriented or optimization AI-oriented using the labeled data. Our best prediction model has an accuracy of 99% in testing data, and the results are consistent with the method of using keywords. For simplicity and interpretability, this study uses the keyword method. Since the dependent variable is product development, there are concerns that startups may invest in AI capabilities in other areas, such as human resource management (e.g., AI-powered recruiting), resulting in errors in our AI measurements. However, more than 95% of AI-related job postings in our sample directly mention keywords such as product development and R&D, suggesting that startups' AI investments are mostly devoted to product innovations in the sample we observe.

***Lean Startup Method (LSM)***

LSM is a process by which firms use to create products. To operationalize the extent to which startups use LSM, we count the number of job posts, social media articles, and news reports that contain keywords related to the lean startup method (Table 1). We follow the previous literature that operationalizes the two major applications of LSM (Ries 2011, Bortolini et al. 2018, Harms and Schwery 2020, Koning et al. 2022) by looking for keywords that represent (1) Prototyping (e.g., trial product) and (2) Controlled Experimentation (e.g., AB test). Startups that hire employees with expertise in experimentation or announce trial product launches are more likely to have adopted LSM than other startups. We then take the first principal component of the factors derived from the three data sources as the firm's **LSM Level** each year. Similar to the AI capability, we found that we cannot reject the hypothesis that all three measures have the same coefficients. For simplicity, we combine them into a single measure. We also separately measure

AI job posting is: "…The algorithm engineer is responsible for using machine learning to optimize R&D processes, for example, image de-noising…". The job title is usually Software Engineer or AI Engineer.

the extent to which companies adopt each of the two major applications of LSM, **Prototyping** and **Controlled Experimentation,** using the keywords.[10] To further ensure that we are measuring the underlying construct, we capture whether the startup has received funding from government or state-owned funds, which can be a proxy for LSM in our context. This is because startups receiving support from public institutions have a greater propensity to produce and follow business plans (Honig and Karlsson 2004), and hence are less likely to adopt the principles of LSM. We also examine the concentration level of investor voting rights as dispersed voting rights are linked to being slower at changing business directions, or lower LSM adoption (Brinckmann et al. 2010). Higher values on these measures represent that the company is more likely to have adopted LSM. The results are listed in Appendix A2 and are consistent with the LSM measurement.

However, firms that do not have any LSM-related job posts, social media articles or news reports can still be using LSM (Type II error). On the other hand, it is less likely that a firm that rates high on our LSM metric will know little about LSM methods (Type I error). Thus, we are more likely to misattribute adopters as non-adopters, which could lead to underestimating the effect of LSM[11] (Griliches and Hausman 1986, Tambe et al. 2012). The same reasoning may apply to the AI measures. Instrumental variable analysis can further alleviate these measurement errors.

***Other Variables***

For each startup in each year, we construct firm-level control variables such as **Age**, **Number of Subsidiaries**, **Number of Branches**, **Total Number of Job Posts** and **Total Number of Employees**, company quality information such as **Total Number of Administrative**

[10] An example of a prototyping job posting is: "…Familiar with Google MVP and other platforms…". The job title is usually Product Engineer. An example of a controlled experimentation job posting is: "…Have an experimentation mindset and familiar with A/B testing…". The job title is usually Data Scientist.

[11] Firms using LSM without mentioning it in job postings, social media articles, or news reports would be mislabeled as non-adopters in our data (Type II error). As a result, the average performance level of LSM non-adopters measured will be higher than their actual performance, given that LSM adopters on average have higher performance levels than non-adopters.

**Approvals**, **Total Number of Patents**, **Total Number of Copyrights** and **Total Number of Certificates**, company funding information such as **Total Funding Rounds**, and **Historical Funding Amounts**. By taking the average number of days between two first-release products and two versions of the same product, we calculate **Average Completion Time for First-Release Products** and **Average Completion Time for Iterative-Release Products** for each company in each year as a proxy for product development speed. We also calculate the **Market Uncertainty** for each company in each year using the failure rate of startups within its industry (Hyytinen et al. 2015). Our unbalanced panel dataset includes 10,641 observations for about 1,800 companies from 2011 to 2020. Table 2 shows summary statistics.[12]

<Table 2 Inserted Here>

## 4. Empirical Methodology

### 4.1 Base Model

We explore the effect of AI capability on the number of product innovations produced in a startup in a typical year. We first fit the following OLS model with company fixed effect ($\gamma_i$) and time dummies ($\tau_t$) for all companies after applying various control variables:

$$\{Number_of_Products_{i,t+1}, Number_of_Novel_Products_{i,t+1}, Number_of_Incremenal_Products_{i,t+1}\} = \beta_0 + \boldsymbol{\beta_1 AI_capability_{i,t}} + controls + \tau_t + \gamma_i + \varepsilon_{i,t}$$

A significant benefit of a firm fixed-effect model is that all the time-invariant characteristics of the startups, such as industry categories and the backgrounds of founders, are controlled in the regression. We lead the dependent variable by one year to avoid potential reverse causality and simultaneity issues. Control variables include firm age, number of branches, number of subsidiaries, number of employees, total trademarks, total patents, total certificates, total copyrights, total funding rounds and amount, and total job posts. This comprehensive set of information about startups can help us control the unobserved effects of "firm quality" that can potentially cause omitted variable biases.

[12] Table of Correlations in Appendix 3.

### 4.2 Complementarity Test

In order to test the complementarities between company AI capability and lean startup practices, we follow the previous literature and conduct two statistical tests: the demand equation that examines the correlations in inputs, and the performance test that examines the performance differences when complementary inputs are used in combination and independently (Arora and Gambardella 1990, Arora 1996, Athey and Stern 1998, Aral and Weill 2007, Brynjolfsson and Milgrom 2013, Aral et al. 2012, Tambe 2014, Wu et al. 2020). Specifically, we fit the following OLS model with company fixed effect ($\gamma_i$) and time dummies ($\tau_t$) to test whether companies that have adopted LSM are more likely to invest in AI capability:

$$AI_capability_{i,t} = \beta_0 + \boldsymbol{\beta_1 LSM_{i,t}} + controls + \tau_t + \gamma_i + \varepsilon_{i,t}$$

Then we fit the following OLS model with company fixed effect ($\gamma_i$) and time dummies ($\tau_t$) to test whether AI capability and LSM form a system of complements that provides additional performance improvements in product development when used together:

$$\{Number_of_Products_{i,t+1}, Number_of_Novel_Products_{i,t+1}, Number_of_Incremental_Products_{i,t+1}\} = \beta_0 + \boldsymbol{\beta_1 AI_capability_{i,t}} + \boldsymbol{\beta_2 LSM_{i,t}} + \boldsymbol{\beta_3 AI_capability_{i,t} \times LSM_{i,t}} + controls + \tau_t + \gamma_i + \varepsilon_{i,t}$$

The complementarities are supported by a positive $\boldsymbol{\beta_3}$ coefficient. One concern about our measurement is that the company's AI capability might be spuriously correlated with its LSM level because they are computed using overlapped sources of information (job posts, social media articles, and news reports). However, the performance test can help rule out such concerns if the estimated coefficient of the interaction term is significantly positive, because if the correlations are spuriously high due to the overlapped sources of data, then most available data will be concentrated in observations with both AI capability and LSM level, while few data in observations with only AI capability but no LSM, or only LSM but no AI, thereby there will not be enough statistical power to yield a statistically significant coefficient $\boldsymbol{\beta_3}$ (Aral et al. 2012).

### 4.3 Identification Strategy

Offering conclusive causal evidence proves challenging due to the endogeneity of organizational practices in observational data. We employ two methods below to alleviate these issues. First, the complementarities framework we use is naturally robust against simple selection and reverse causality biases. Second, we use AI policy changes in China that provide some exogenous variation in AI investments. We also use neighboring firms as an instrument for one's own AI and LSM adoption.

#### 4.3.1 Identification through Complementarities

Complementarity approaches are naturally robust to some types of endogeneity and reverse causality problems because they are about matching two or more investment decisions rather than about the effectiveness of the decisions themselves. Any biases that affect the complementarity term must be present only at the confluence of both factors, and not when factors are present individually (Tambe et al. 2012). In addition, the two tests for complementarities are subject to different sources of bias, so consistent results provide greater confidence that the results are not primarily due to endogenous issues (Wu et al. 2020).

#### 4.3.2 Instrumental Variables for AI

Nonetheless, the decision to adopt AI and LSM can be endogenous due to unobservable characteristics such as startup quality. Such issues are partially addressed using a firm fixed effect model and the detailed company quality information in each year as control variables. In addition, we use instrumental variables to address other forms of endogeneity. We use two sets of instruments for AI capabilities. The first set of instrumental variables includes all the AI policies issued by provincial-level governments in China from 2011 to 2020 collected from the official government websites. Over the past decade, many local governments in China have offered incentives, such as cash subsidies and tax abatements[13], to encourage local businesses to adopt AI technologies. These incentives are often unexpected, making them an exogenous source of variation useful for studying company performance. Such preferential policies are critical for startups as the tax benefits they receive after adopting AI can partially offset the investment costs associated with AI implementation. Thus, we create an instrumental variable for a firm's AI investment using the cumulative number of active AI policies in the province where the company is located, measured at the end of each year. We identified a total of 3,231 policies containing AI-related keywords from official government websites in China. Figure 2 shows the provincial distribution of the total number of AI incentives in China from 2011 to 2020.

<Figure 2 Inserted Here>

This instrumental variable varies at the region-year level. One-way ANOVA test of the variable shows that cross-region variation accounts for 45% of total variation, while within-region cross-time variation accounts for 55% (details are in Appendix A4). The geographic distribution of the sample startups is as follows: 25% are in Beijing, 15% in Shanghai, 15% in Guangdong, with the remaining startups distributed relatively evenly across other regions of China. This distribution offers sufficient variation to provide the necessary power for the first-stage analysis of the 2SLS model.

The second set of instrumental variables is the number of news reports each year about companies in the same industry as the focal startup adopting AI. These reports are likely exogenous to the focal startup, as news of a competitor adopting new technology can prompt imitation

[13] An example of these policies: "…R&D expenses in core tech domains such as artificial intelligence can be deducted at a rate of 175%…".

(Gatignon and Robertson 1989, Biemans 2018). This instrumental variable varies at the industry-year level. Where possible, we use both sets of instrumental variables to generate exogenous variations for discovery- and optimization-oriented AI capabilities.

A concern with the instrumental variables is that AI policies published by local governments might correlate with regional socio-economic levels and, consequently, with unobserved firm qualities. Similarly, the number of AI-related news items from companies in the same industry could be correlated with unobserved market dynamics and firm qualities. If this were the case, we would expect these instrumental variables to be associated with startup innovation activities even for startups that have not adopted AI.

To address this, we follow the procedures in Martin and Yurukoglu (2017) and Wang et al. (2023) and conduct a reduced-form instrumental variable test, with results reported in Appendix A5. Specifically, we create a binary variable for AI adoption, identifying the year a company first lists AI-related job postings, publishes AI-related social media articles, or is mentioned in a news report. We then separate our sample into AI adopters and non-adopters and use the instrumental variables as the independent variable. The results show that our instrumental variables are only associated with startup innovation activities when the startups have adopted AI, providing further validation of the instrumental variables.

**4.3.3. Instrumental variables for LSM**

We also try to explore potential instrumental variables for LSM. Specifically, we use the average level of prototyping and controlled experimentation of other companies[14] each year that (1) share a board member with the focal startup and (2) are in a different industry and province from the focal startup. Companies with a shared board member are likely to receive similar advice and consequently likely to formulate similar managerial and operational strategies (Kroll et al. 2008, Dechow and Tan 2021). At the same time, they are less likely to experience the same systematic shocks in performance faced by firms in the same industry or region. Thus, their product innovation performances are less likely to be correlated. Where possible, we use both instrumental variables to generate exogenous variations for prototyping and controlled experimentations. The first-stage F-statistics for all 2SLS results are over 20, suggesting that the results do not suffer from weak instrument issues.

[14] Including companies that are outside our sample.

Although our instrumental variables and the use of the complementarity framework may not effectively address all endogeneity issues, their combined application should significantly diminish potential biases in our results stemming from causality problems. Nevertheless, it is important to acknowledge the limitations of our archival data, as it may not definitively address causality, especially when exploring the economic impact across a broad sector of the economy, as is the focus of our study.

## 5. Results and Interpretation

We first examine the effect of AI adoption on startups' product innovations in Table 3. Columns 1 to 6 show the effect of AI on software startups. We show that AI capability is positively associated with increased future software copyrights, including both first and iterative releases, after controlling for various firm characteristics, and 2SLS results are similar.[15] Economically, the results show that a one-standard-deviation increase in AI capability, which includes about 30 AI-related job postings, 50 AI-related social media articles, and 15 AI-related news reports, is associated with creating about 0.6 first-release software product and 0.4 iterative-release software product in the following year. Since the average number of new software copyrights every year for software companies is only 9, this increase is substantial. To show that this effect is not only applicable to digital products, Columns 7 to 12 show the positive effect of AI capability for firms that produce physical products. Specifically, a one-standard-deviation increase of AI capability is associated with creating about 2.5 first-release products and 2 iterative-release products in the following year. As the average number of new product trademarks every year for physical product companies is about 50, this means a 10% increase. These results support H1.

[15] Due to the significant increases in the coefficients in the 2SLS regressions we restrict our discussion to the sign of these coefficients, although these could suggest the magnitude of the effects we observe are in reality larger than what is suggested by the OLS results.

<Table 3 Inserted Here>

We then examine the complementarity between AI capability and LSM level by conducting an event study analysis that compares the effect of AI adoption on product innovation performance in startups that have adopted some level of LSM versus those that have not (Figure 3). We categorize the sample startups into LSM adopters and non-adopters based on whether the company has ever published an LSM-related job posting, social media article, or news report. Each group comprises at least 30% of the total observations, providing sufficient statistical power for identification.

To account for potential biases caused by different timings of AI adoption in our panel data, we use the doubly robust difference-in-difference estimator (Callaway and Sant'Anna 2021; Sant'Anna and Zhao 2020), following Goodman-Bacon's (2021) guidelines. AI adoption is defined as the first instance of an AI-related job posting, social media article, or news report. The top two plots in Figure 3 show the effect of AI adoption on software startups: the left plot is for firms that did not adopt LSM, and the right plot is for LSM adopters. The results indicate that AI adoption has a limited effect on non-adopters but a significant effect on LSM adopters, with a one-year lag likely due to the time required to develop a new product. These findings not only confirm the positive impact of AI adoption on startup product innovation but also support the complementarity between AI capability and LSM. The bottom two plots in Figure 3 show similar results for physical product startups. It is worth noting that for physical products, adopting AI without implementing LSM may even negatively impact innovation. AI investments are costly, and using AI in areas it is not designed to support can be detrimental to firms.

<Figure 3 Inserted Here>

The regression results are reported in Table 4. We first show the correlation test results in Column 1 for software and Column 6 for physical product companies. The estimated coefficient is positive. Next, we test the interaction effects between AI capability and LSM level in the performance regression. For software companies, Columns 2 and 3 show that the adoption of both AI and LSM as a system is significantly associated with producing more future software products. Economically, the results show that a one-standard-deviation increase in AI capability is associated with 1.8 more software copyrights in firms that are one standard deviation above the mean in LSM adoption, which is a 20% increase in the number of products, given that an average startup develops 9 software copyrights in the sample. Both the correlation and performance tests suggest a complementary relationship between AI capability and LSM. The 2SLS results are larger in magnitude but directionally consistent. Note that after adding the interaction term between AI capability and LSM, the coefficient of AI capability alone becomes negative, suggesting that only increasing AI capability without LSM can negatively impact the innovation performance in companies; however, if increasing both AI capability and LSM together, the overall effect is positive.[16] Columns 4 and 5 show that the adoption of both AI and LSM as a system is also significantly associated with reducing product completion time. Economically, the results show that a one-standard-deviation increase in AI capability is associated with a reduction of 45 days in product development time in firms that are one standard deviation above the mean in LSM adoption, which is a 25% reduction in development time, given that an average startup takes 180 days to develop a new product in the sample. The results for physical products (columns 6-10) largely mirror the results for digital products. Collectively, they indicate that complementarity exists between AI capability and LSM for developing both software and physical products. All these results support H2.

[16] The effect of 1 standard-deviation increase in AI capability alone is -2.694, while the effect of 1 standard-deviation increase in both AI capability and LSM level is -2.694+0.998+1.791=0.095, which is positive.

<Table 4 Inserted Here>

In Table 5, we explore the interplay between different facets of AI capability (discovery-oriented AI versus optimization-oriented AI) and different facets of LSM (prototyping and controlled experimentation). Column 1 shows that prototyping and discovery-oriented AI are complements in developing first-release software, while the coefficients of the other three interactions are either not statistically significant or negative. Economically, a one-standard-deviation increase in both discovery-oriented AI and prototyping is associated with an increase of about 0.4 first-release software copyright. Column 2 uses instrumental variable 2SLS to enhance identification, and the results are consistent. Columns 3 and 4, on the contrary, show that controlled experimentation and optimization-oriented AI are complements in developing iterative-release software copyrights, while the coefficients of the other three interactions are either not statistically significant or negative. Economically, a one-standard-deviation increase in both discovery-oriented AI and controlled experimentation is associated with an increase of about 0.5 iterative-release software copyright. Columns 5 to 8 show the results for physical products. They are qualitatively similar to what we found for software.

<Table 5 Inserted Here>

Collectively, there are two implications suggested by these results. First, columns 1, 2, 5 and 6 support the mechanism that discovery-oriented AI and prototyping are complements for developing first releases and expanding into new markets. Columns 3, 4, 7 and 8 support the mechanism that optimization-oriented AI and controlled experimentation are also complements for fast iterations, supporting H3a and H3b. Second, the coefficients for the cross complementarities (i.e., discovery-oriented AI * controlled experimentation and optimization-oriented AI * prototyping) are either statistically insignificant or negative, which verifies the theoretical analysis above and also suggests that misallocation of complementary resources can potentially reduce innovation (Aral et al. 2012, Tambe 2014). These findings also rule out alternative explanations, such as the presence of slack resources, because the cross complementarities would be expected to exhibit significance in that case as well.

## 6. Mechanism and Robustness Check

### 6.1. Discovery-oriented AI and Prototyping to Reduce Market Uncertainty

If the complementarities between discovery-oriented AI and prototyping help firms innovate when searching for and validating a niche market is difficult, this effect should be stronger in environments with high market uncertainty, where such tasks are especially challenging. As described in the data section, we measure uncertainity at the industry level by calculating the failure rate of startups in its industry. Table 6 provides this evidence, showing that the three-way complementarity between prototyping, discovery-oriented AI, and market uncertainty is positive. Economically, a one-standard-deviation increase in both prototyping and discovery-oriented AI capability is associated with an additional 1.7 first-release software copyrights and 2.3 first-release product trademarks in markets with uncertainty levels one standard deviation above the mean. This finding suggests that the complementarity between discovery-oriented AI and prototyping in developing first releases is stronger when startups face higher market uncertainty, supporting the proposed mechanism.

<Table 6 Inserted Here>

### 6.2. Reducing Development Time Through AI-LSM Complementarity

We find that combining discovery-oriented AI with prototyping significantly reduces development time for first-release products, while pairing optimization-oriented AI with controlled experimentation shortens development time for iterative-release products. To examine this, we use the average completion time for first-release and iterative-release products as dependent variables. Table 7 presents the results: Columns (1)–(2) show estimates for first-release products, and Columns (3)–(4) for iterative-release products. These findings are consistent with those in Table 5 and reinforce the mechanism of complementarity—discovery-oriented AI complements prototyping in accelerating early-stage product development, while optimization-oriented AI complements controlled experimentation in refining existing products. Furthermore, these results extend those in Table 5 by showing that startups not only develop more products but also complete them more quickly. The complementarity between AI and the Lean Startup Methodology is especially pronounced for first-release products: with an average development time of 170 days for new products and 220 days for iterative ones, the AI-LSM combinations reduce development time by approximately 15% and 5%, respectively.

<Table 7 Inserted Here>

### 6.3. Complementarity in Market Expansion and Product Quality

In China, companies are legally permitted to operate only within the business domains listed on their business licenses. To enter a new market niche, firms must formally apply to add a new business scope entry. Accordingly, we measure business scope expansion by counting the number of new entries added to a company's license each year. In Columns (5) and (6), we show that the joint adoption of discovery-oriented AI and prototyping is associated with broader business scope expansion. Given that the average firm adds about two new business entries annually, this represents a 2% to 7% increase. These results are consistent with those in Table 5, reinforcing the complementarity between discovery-oriented AI and prototyping in enabling startups to explore and enter new markets.

Product quality is also examined in Columns (5) and (6), measured by the number of certifications granted by neutral third parties, such as standard-setting organizations. We find that the combined use of optimization-oriented AI and controlled experimentation is associated with a higher number of certified, high-quality products. Specifically, a one-standard-deviation increase in both capabilities leads to an increase of 0.15 high-quality products. These findings further support the complementarity between optimization-oriented AI and controlled experimentation in enhancing product refinement and quality.

## 7. Discussion

Our findings highlight the importance of treating AI as a heterogeneous construct, as different AI capabilities require distinct organizational processes for effective use. We provide new evidence that the impact of AI on product development is strengthened through its complementarity with LSM. Joint adoption of AI and LSM is associated with increased product innovation—both in first releases and iterative improvements—as well as faster development cycles, higher product quality, and broader business scope, particularly in uncertain industries. By contrast, adopting AI without LSM yields limited gains. Further analysis shows that discovery-oriented AI complements prototyping in supporting first releases and market expansion, while optimization-oriented AI complements controlled experimentation in refining product iterations and quality.

Although AI adoption has surged in recent years, many firms have yet to realize meaningful returns. Our findings offer a possible explanation: performance gains depend on aligning specific AI capabilities with organizational practices like LSM. While our study focuses on China—a global AI leader—we believe the results generalize more broadly, as the tools and innovation processes are globally relevant. These insights also apply to established firms engaging in intrapreneurship, not just startups.

This study has limitations. In novel or data-scarce domains, firms face higher uncertainty and may require multiple cycles of trial and error with AI and LSM before achieving product success. This highlights the need for future research into how AI can support learning under data scarcity, especially within the experimentation-feedback loop of product development.

Nonetheless, this work offers actionable insights. For startups, while modern AI tools offer predictive power, their effectiveness depends on high-quality data and interpretability. Iterative market feedback is essential for refining AI applications. Coupling AI with LSM helps firms adjust what the AI optimizes for and validate outputs through real-world experimentation. For investors and policymakers, our results emphasize the importance of organizational complementarity—matching tools to objectives and building capacity around them. Simply encouraging AI adoption, without regard to how it is implemented, may yield limited or even negative returns.

This paper contributes to research on the business value of AI and its organizational complements. We show that aligning AI capabilities with specific objectives and complementary practices like LSM can help explain the uneven returns observed among early adopters. Startups in particular can benefit from matching the right AI tools to the right stages of product development.

Doing so enables more scalable and effective innovation.

Future research should explore other forms of complementarity, such as workforce skills and organizational structures. As AI continues to evolve, assessing the business value of tools like large generative models and interpretable machine learning—and identifying the organizational conditions needed to deploy them effectively—will be increasingly important. Our results point to the need for more precise understanding of what makes AI work in practice.

**Figure 1: An Example of Software Copyright and Trademark (Translated in English)**

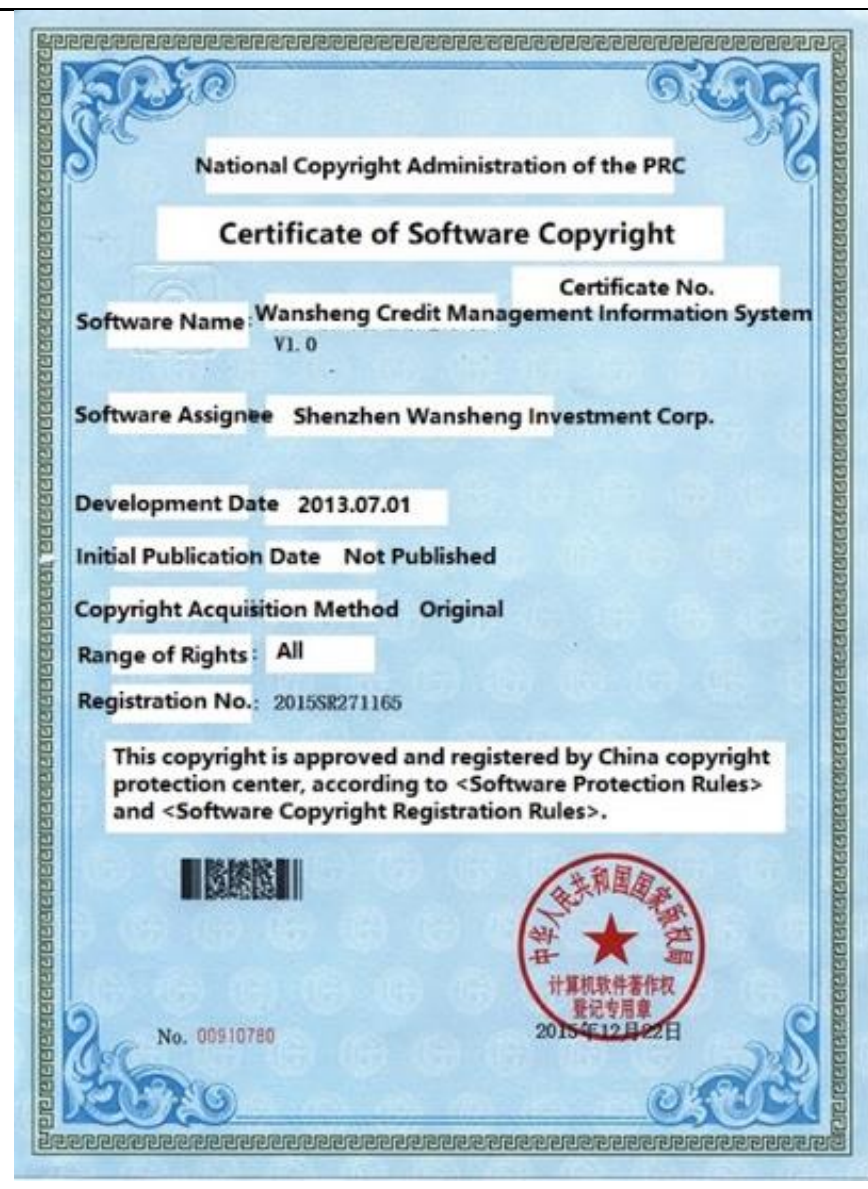

Software Copyright (Example)

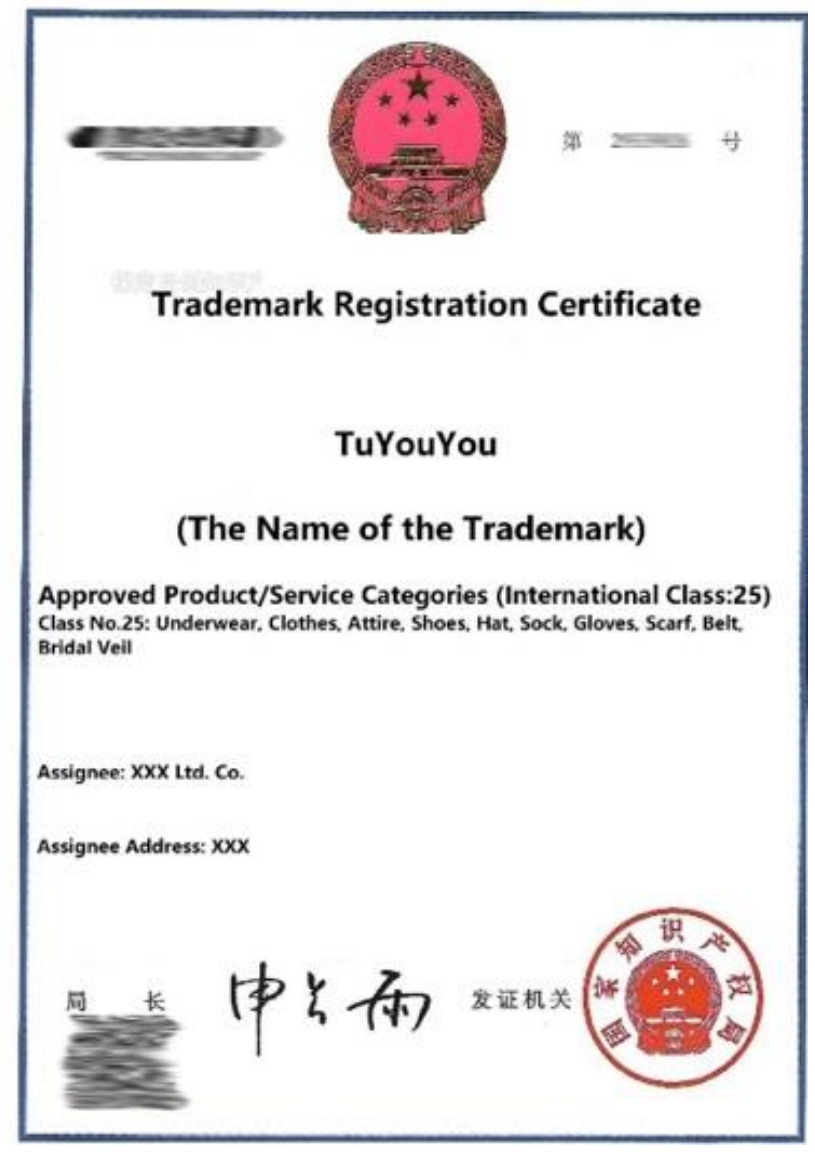

Trademark (Example)

**Figure 2: Provincial Distributions of AI Incentives in China (2011-2020)**

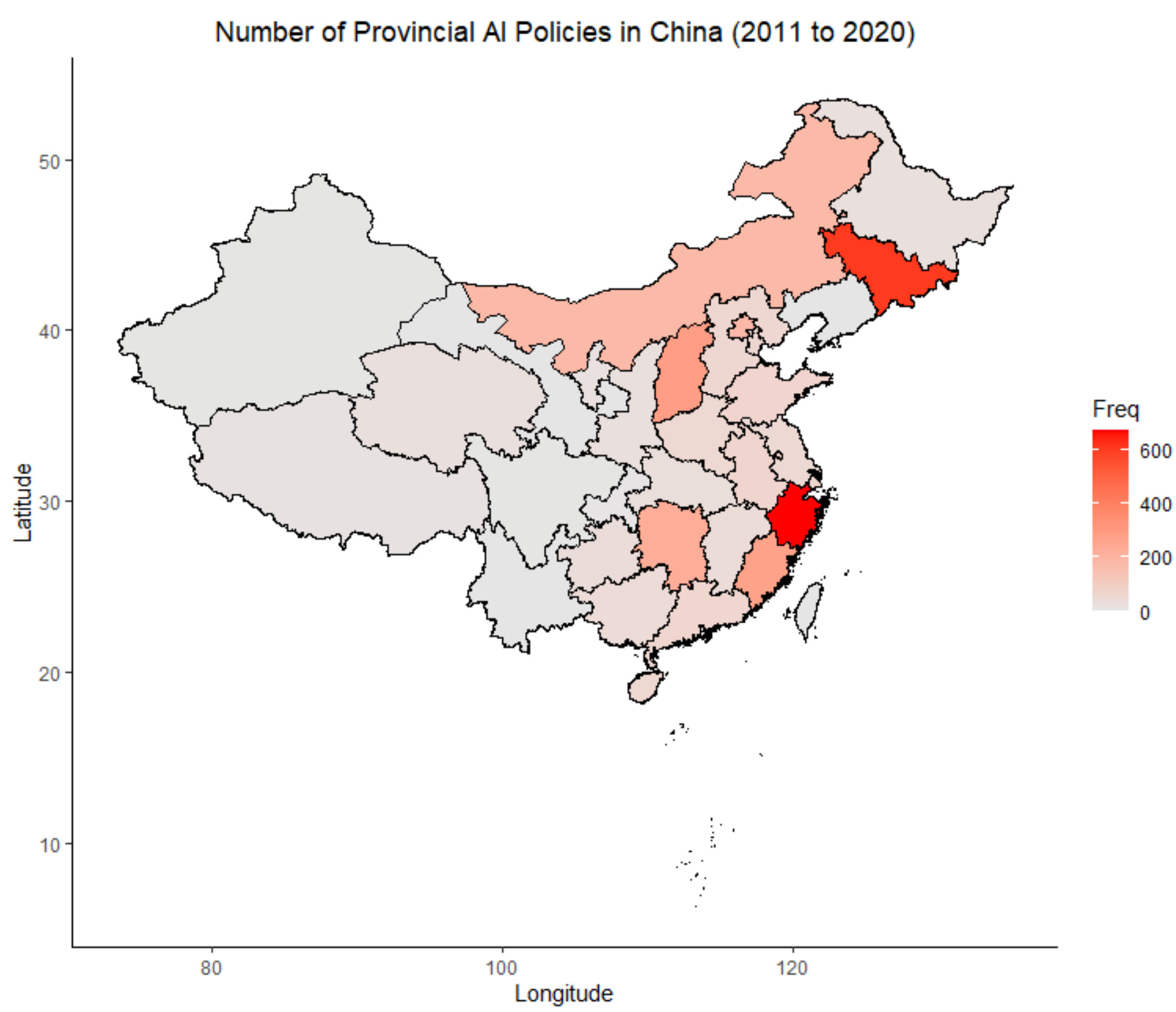

**Figure 3: Event Study: Effect of AI Adoption on Product Innovation based on Callaway and Sant'Anna (2021) and Sant'Anna and Zhao (2020)**

| | Non LSM Adopters | LSM Adopters |
|---|---|---|
| **Software Startups (DV: Lead Number of Software Copyrights)** | 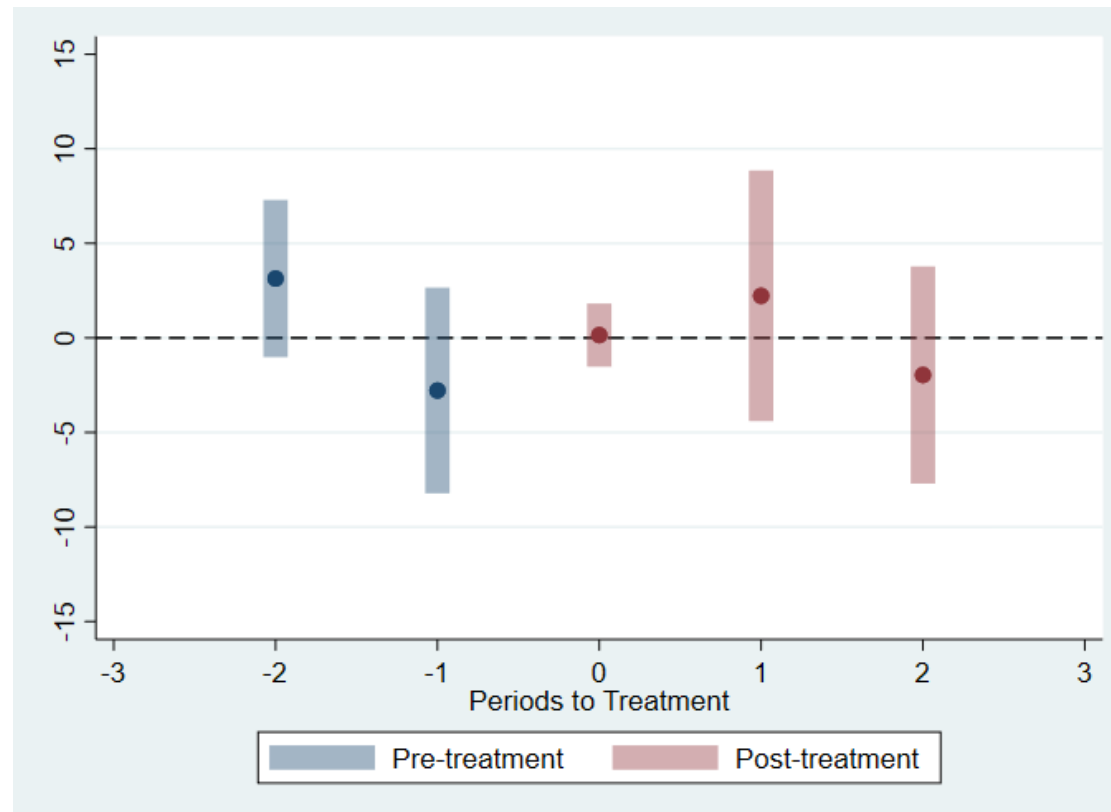 | 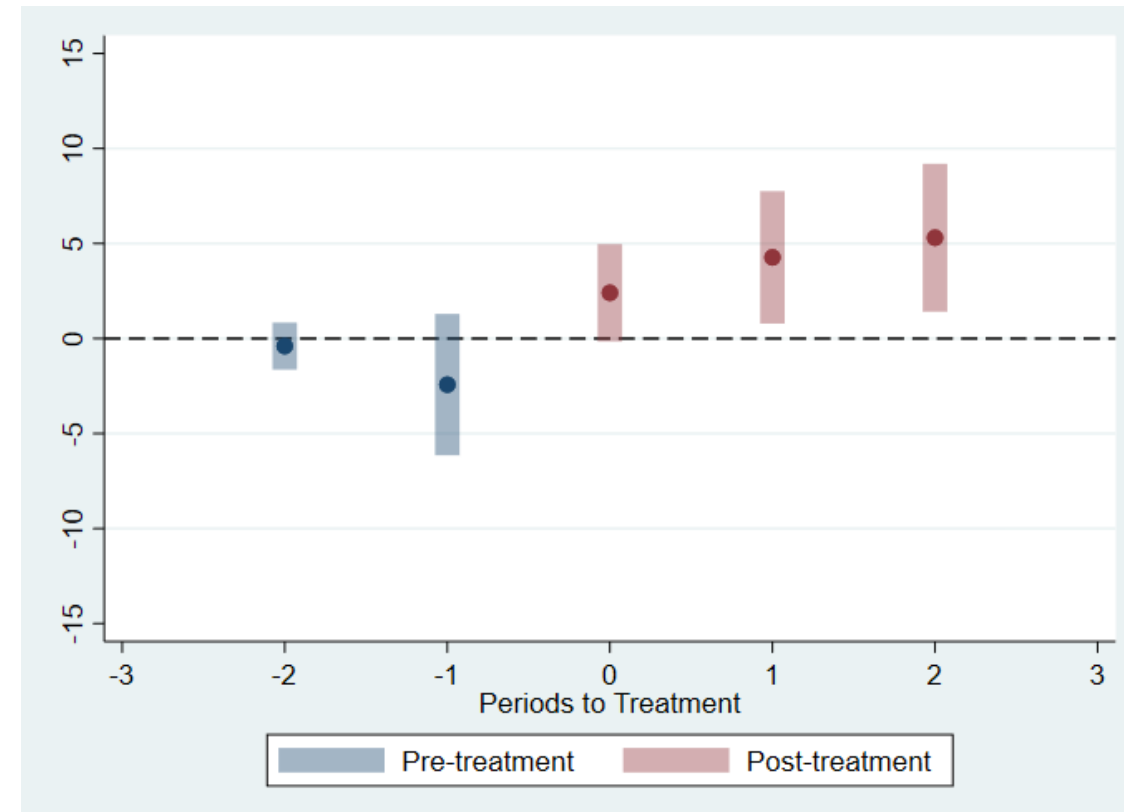 |
| **Physical Product Startups (DV: Lead Number of Product Trademarks)** | 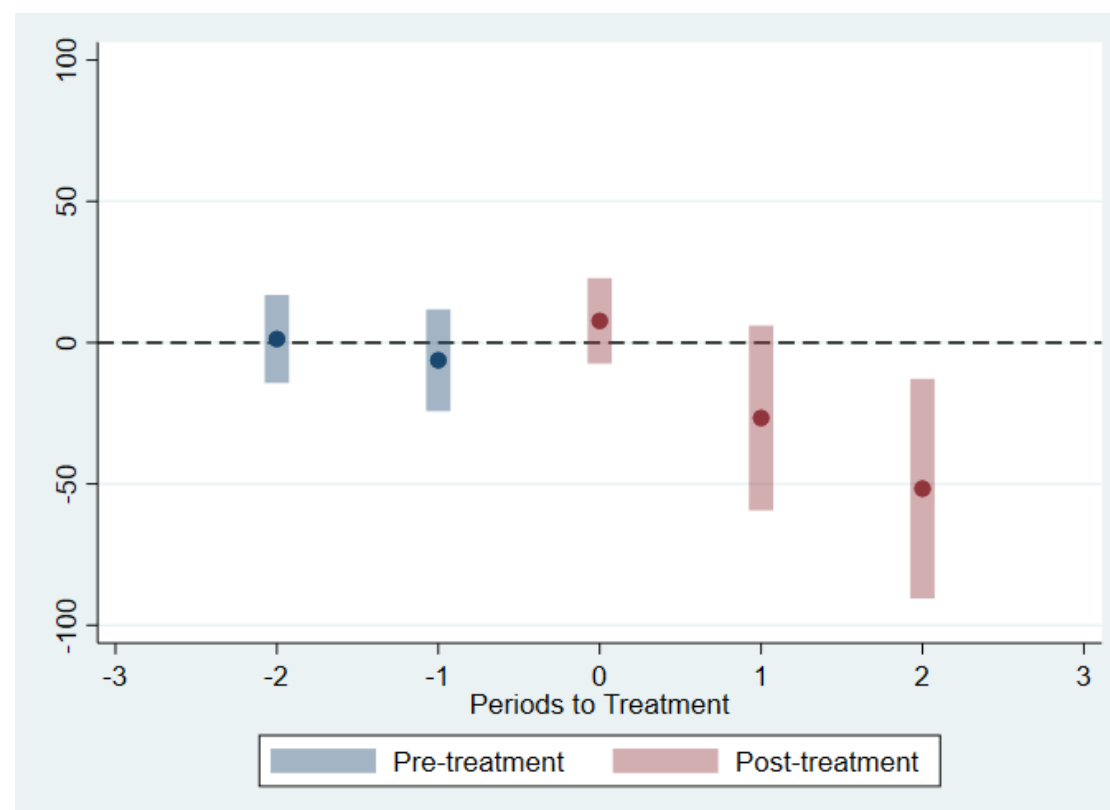 | 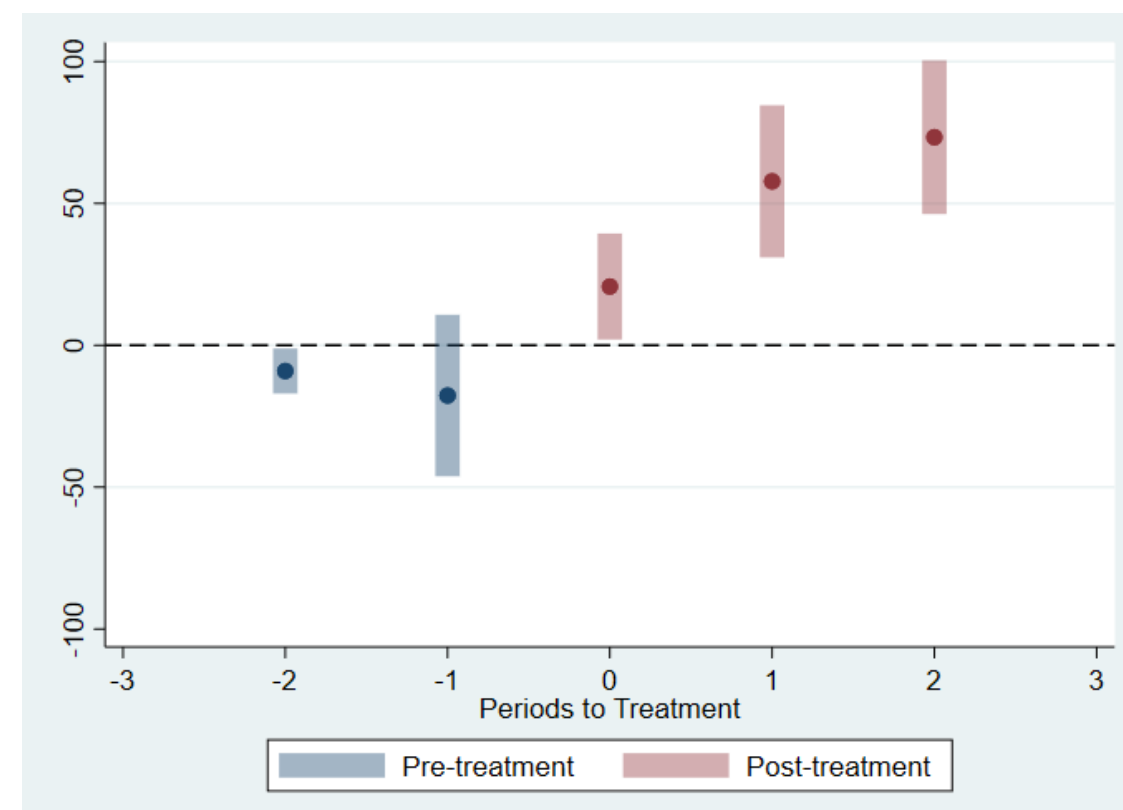 |

Notes: Event study plots created using *csdid* package in STATA. Top 2 plots show AI adoption effect on software startups. Bottom 2 plots show AI adoption effect on physical product startups. Left 2 plots are companies that never adopted LSM. Right 2 plots are companies that have at least one job posting, social media article or news report that is LSM-related.

**Table 1: Key Words Dictionary**

| Topic | Keywords | Translation in Chinese |
|---|---|---|
| AI | Artificial Intelligence, AI, ML (Machine Learning), DL (Deep Leaning), CV(Computer Vision), Neural Network, CNN, RNN, NLP (Natural Language Processing), Supervised Learning, Unsupervised Learning, XGBoost, Tensorflow, ND4J, (LSTM) Long Short-Term Memory, Keras | 人工智能，AI，ML（机器学习）[17]，DL（深度学习），CV（计算机视觉），神经网络，CNN，RNN，监督学习，无监督学习，XGBoost，Tensorflow，ND4J，LSTM，Keras |
| Optimization-oriented | Refine, Efficiency, Exploitation, Optimize, Automate, Process, Tune | 精炼，效率，开发利用，优化，自动化，流程，微调 |
| Discovery-oriented | Explore, Search, Discover, Insight, Analyze, Forecast, Detect | 探索，搜索，寻找，发现，洞见，分析，预测，识别 |
| Lean Startup Method (LSM) | MVP (Minimum Viable Product), Test Product, Experimentation, Lab (Technician), (A/B) Testing, Validation, Verification, Trial and Error, Customer/Market Demand, Customer/Market Learning/Response | MVP（最简可行产品），试产品，试验，实验室/员，(A/B)测试，验证，检验，顾客/市场需求，顾客/市场学习/响应 |
| Prototyping | MVP (Minimum Viable Product), Test Product, Validation, Verification, Trial and Error, Customer/Market Demand, Customer/Market Learning/Response | MVP（最简可行产品），试产品，验证，检验，试错，顾客/市场需求，顾客/市场学习/响应 |
| Controlled Experimentation | Experimentation, Lab (Technician), A/B Testing | 试验/实验，实验室/员，A/B 测试 |

**Table 2: Summary Statistics**

| Panel A: Cross-Section Statistics | | |
|---|---|---|
| | **Software Companies** | **Physical Product Companies** |
| Share of Firms ever Adopting AI | 0.75 | 0.73 |
| Share of Firms ever Adopting Discovery-oriented AI | 0.56 | 0.59 |
| Share of Firms ever Adopting Optimization-oriented AI | 0.58 | 0.57 |
| Share of Firms ever Adopting LSM | 0.74 | 0.72 |
| Share of Firms ever Adopting Prototyping | 0.56 | 0.54 |

[17] We also manually checked and dropped keywords that have “AI” or “ML” in them while having nothing to do with artificial intelligence or machine learning (such as “AIC” or “HTML”).

| | | |
|---|---|---|
| Share of Firms ever Adopting Controlled Experimentation | 0.69 | 0.67 |

| Panel B: Panel-Data Statistics | | | | | | | | | | | | |
|---|---|---|---|---|---|---|---|---|---|---|---|---|
| | Software Companies | | | | | | Physical Product Companies | | | | | |
| **Variable Name** | **No. Obs.** | **Mean** | **Std. Dev.** | **Min** | **Median** | **Max** | **No. Obs.** | **Mean** | **Std. Dev.** | **Min** | **Median** | **Max** |
| Lead Number of Software Copyrights | 2,962 | 9.00 | 8.68 | 0 | 7 | 129 | 7,679 | 13.02 | 7.66 | 0 | 12 | 100 |
| Lead Number of Trademarks | 2,962 | 50.44 | 149.72 | 0 | 34 | 751 | 7,679 | 59.53 | 77.50 | 0 | 44 | 2416 |
| Lead Number of First-Release Software Copyrights | 2,962 | 4.87 | 6.61 | 0 | 5 | 113 | 7,679 | 4.32 | 4.89 | 0 | 5 | 92 |
| Lead Number of First-Release Trademarks | 2,962 | 20.33 | 32.55 | 0 | 10 | 565 | 7,679 | 22.11 | 51.50 | 0 | 11 | 2157 |
| Lead Number of Iterative-Release Software Copyrights | 2,962 | 4.13 | 4.32 | 0 | 6 | 59 | 7,679 | 4.03 | 3.99 | 0 | 7 | 39 |
| Lead Number of Iterative-Release Trademarks | 2,962 | 30.11 | 28.39 | 0 | 23 | 386 | 7,679 | 34.60 | 39.85 | 0 | 23 | 834 |
| Lead Number of High-Quality Products | 2,962 | 1.76 | 3.44 | 0 | 2 | 103 | 7,679 | 2.36 | 4.23 | 0 | 3 | 170 |
| Lead Average Completion Time (in Days) for First-Release Products | 2,962 | 174.10 | 145.97 | 3.23 | 73 | 360 | 7,679 | 177.07 | 145.66 | 3.97 | 73 | 360 |
| Lead Average Completion Time (in Days) for Iterative-Release Products | 2,962 | 226.60 | 122.37 | 1.00 | 190 | 360 | 7,679 | 214.20 | 124.10 | 0.47 | 150 | 360 |
| Lead # New Business Scope Entries | 2,962 | 2.76 | 5.29 | 0 | 1 | 105 | 7,679 | 2.11 | 3.37 | 0 | 1 | 89 |
| AI Capability | 2,962 | -0.11 | 0.95 | -0.96 | -0.27 | 14.60 | 7,679 | 0.04 | 1.02 | -0.96 | -0.08 | 14.34 |
| Discovery-oriented AI | 2,962 | -0.12 | 1.34 | -1.07 | -0.85 | 18.55 | 7,679 | 0.04 | 1.32 | -1.07 | -0.16 | 15.22 |
| Optimization-oriented AI | 2,962 | -0.21 | 1.14 | -1.06 | -0.59 | 20.76 | 7,679 | 0.08 | 1.36 | -1.06 | -0.30 | 22.32 |
| LSM Level | 2,962 | -0.03 | 0.96 | -1.17 | -0.23 | 13.20 | 7,679 | 0.01 | 1.01 | -1.17 | -0.17 | 12.57 |
| Prototyping | 2,962 | 0.03 | 1.49 | -1.32 | -0.16 | 19.14 | 7,679 | -0.01 | 1.42 | -1.32 | -0.40 | 16.95 |
| Controlled Experimentation | 2,962 | 0 | 1.24 | -1.06 | -0.17 | 15.86 | 7,679 | 0 | 1.24 | -1.06 | -0.08 | 16.82 |
| Age | 2,962 | 3.83 | 2.10 | 1 | 4 | 9 | 7,679 | 3.84 | 2.10 | 1 | 4 | 9 |
| Number of Branches | 2,962 | 3.74 | 23.39 | 0 | 0 | 367 | 7,679 | 2.71 | 15.21 | 0 | 0 | 224 |
| Number of Subsidiaries | 2,962 | 4.94 | 7.94 | 0 | 3 | 71 | 7,679 | 5.03 | 8.07 | 0 | 2 | 82 |
| Total Number of Job Posts | 2,962 | 112.57 | 350.11 | 0 | 220 | 20555 | 7,679 | 166.55 | 339.07 | 0 | 169 | 44938 |
| Total Number of Employees | 2,962 | 132.19 | 638.63 | 0 | 33 | 24091 | 7,679 | 134.53 | 623.25 | 0 | 34 | 35935 |
| Market Uncertainty | 2,962 | 0.67 | 0.07 | 0.53 | 0.70 | 0.72 | 7,679 | 0.64 | 0.14 | 0.23 | 0.71 | 0.72 |
| Total Number of Administrative Approvals | 2,962 | 2.03 | 4.22 | 0 | 1 | 96 | 7,679 | 1.67 | 2.70 | 0 | 1 | 74 |

| | | | | | | | | | | | |
|---|---|---|---|---|---|---|---|---|---|---|---|
| Total Number of Patents | 2,962 | 26.48 | 164.47 | 0 | 0 | 4134 | 7,679 | 22.22 | 111.34 | 0 | 0 | 2232 |
| Total Number of Certificates | 2,962 | 1.38 | 6.30 | 0 | 0 | 171 | 7,679 | 1.62 | 10.44 | 0 | 0 | 553 |
| Total Number of Work Copyrights | 2,962 | 5.21 | 58.27 | 0 | 0 | 47611 | 7,679 | 13.38 | 665.67 | 0 | 0 | 47611 |
| Total Funding Rounds | 2,962 | 1.34 | 1.32 | 0 | 1 | 8 | 7,679 | 1.32 | 1.42 | 0 | 1 | 17 |
| Total Funding Amount | 2,962 | 7.8e7 | 7.5e8 | 0 | 2.0e6 | 5.0e9 | 7,679 | 6.2e7 | 5.3e8 | 0 | 2.8e6 | 4.8e9 |

## Table 3: AI Capability Effect on Startup Product Development

| Company Industry | Software Companies | | | | | | Physical Product Companies | | | | | |
|---|---|---|---|---|---|---|---|---|---|---|---|---|
| | (1) | (2) | (3) | (4) | (5) | (6) | (7) | (8) | (9) | (10) | (11) | (12) |
| Dependent Variables | Lead # Software Copyright | | Lead # First-Release Software Copyright | | Lead # Iterative-Release Software Copyright | | Lead # Product Trademark | | Lead # First-Release Product Trademark | | Lead # Iterative-Release Product Trademark | |
| Model | FE | FE+2SLS | FE | FE+2SLS | FE | FE+2SLS | FE | FE+2SLS | FE | FE+2SLS | FE | FE+2SLS |
| AI Capability | 1.02*** | **1.99***** | 0.59** | **1.33***** | 0.43*** | **0.63**** | 4.32*** | **20.3***** | 2.49** | **10.1***** | 1.83** | **18.8***** |
| | (0.30) | **(0.49)** | (0.24) | **(0.42)** | (0.14) | **(0.26)** | (1.84) | **(4.68)** | (1.28) | **(3.23)** | (0.86) | **(2.72)** |
| Controls | Firm Fixed Effect, Year Fixed Effect, Company Age, Total Number of Employees, Total Number of Job Postings, Total Number of Patents, Total Number of Certificates, Total Number of Work Copyrights, Total Number of Administrative Approvals, Total Amount of Funding | | | | | | | | | | | |
| Observations | 2,962 | 2,962 | 2,962 | 2,962 | 2,962 | 2,962 | 7,679 | 7,679 | 7,679 | 7,679 | 7,679 | 7,679 |
| R-squared | 0.386 | \ | 0.346 | \ | 0.548 | \ | 0.466 | \ | 0.429 | \ | 0.343 | \ |
| 1st stage F-stat | \ | 645.06 | \ | 645.06 | \ | 645.06 | \ | 602.12 | \ | 602.12 | \ | 602.12 |

Notes: 1. Dependent variable is the number of new products (total number, first-release product, iterative-release product) in the next year. AI Capability is standardized, where one-standard-deviation change is associated with about 30 AI job postings, 50 AI social media articles and 15 AI news reports.

2. Odd number columns report the OLS results with company fixed effect. Even number columns report the 2SLS results with company fixed effect. 2 sets of instruments: (1) Number of AI policies issued by local governments in each year; (2) Number of news reports that a company in the same industry has adopted AI.

3. Robust standard errors in parentheses: *** p<0.01, ** p<0.05, * p<0.1. Bold entries show 2SLS estimates. Statistics that are not applicable are shown in "\".

## Table 4: Tests for Complementarity between AI Capability and Lean Startup Method (LSM)

| Company Industry | Software Companies | | | | | Physical Product Companies | | | | |
|---|---|---|---|---|---|---|---|---|---|---|
| | (1) | (2) | (3) | (4) | (5) | (6) | (7) | (8) | (9) | (10) |
| Dependent Variables | AI Capability | Lead # Software Copyright | | Lead Average Completion Time (in Days) for Products | | AI Capability | Lead # Product Trademark | | Lead Average Completion Time (in Days) for Products | |
| Model | FE | FE | FE+2SLS | FE | FE+2SLS | FE | FE | FE+2SLS | FE | FE+2SLS |
| AI Capability | | -2.694*** | -10.62*** | 80.48*** | 417.74*** | | -22.85*** | -52.68*** | 49.74*** | 125.76*** |
| | | (0.684) | (2.780) | (12.25) | (57.25) | | (2.889) | (13.88) | (6.324) | (30.56) |
| LSM | 0.409*** | 0.998*** | 3.291*** | -27.66*** | -125.30*** | 0.381*** | 5.887*** | 15.08*** | 12.67*** | -10.77 |
| | (0.064) | (0.339) | (0.853) | (6.069) | (17.58) | (0.008) | (1.697) | (4.52) | (3.715) | (9.960) |
| AI * LSM | | 1.791*** | **5.633*** ** | -45.78*** | **-209.36*** ** | | 22.77*** | **45.30*** ** | -72.10*** | **-129.52*** ** |
| | | (0.283) | **(1.336)** | (5.071) | **(27.52)** | | (1.509) | **(10.36)** | (3.302) | **(22.82)** |
| Controls | Firm Fixed Effect, Year Fixed Effect, Company Age, Total Number of Employees, Total Number of Job Postings, Total Number of Patents, Total Number of Certificates, Total Number of Work Copyrights, Total Number of Administrative Approvals, Total Amount of Funding | | | | | | | | | |
| Observations | 2,962 | 2,962 | 2,962 | 2,962 | 2,962 | 7,679 | 7,679 | 7,679 | 7,679 | 7,679 |
| R-squared | 0.595 | 0.485 | \ | 0.315 | \ | 0.667 | 0.591 | \ | 0.319 | \ |
| 1st stage F-stat | \ | \ | 62.19 | \ | 62.19 | \ | \ | 72.07 | \ | 72.07 |

Notes: 1. For Column (1) and (6), dependent variable is AI Capability in the current year. For Column (2)-(3) and (7)-(8), dependent variable is the number of new products (total number) in the next year. For Column (4)-(5) and (9)-(10), dependent variable is the average completion time (in days) for products in the next year.

2. AI Capability is standardized, where one-standard-deviation change is associated with about 30 AI job postings, 50 AI social media articles and 15 AI news reports.

3. Column (5) and (10) report the 2SLS results with both AI Capability and LSM instrumented.

4. Robust standard errors in parentheses: *** p<0.01, ** p<0.05, * p<0.1. Bold entries show 2SLS estimates. Statistics that are not applicable are shown in "\".

## Table 5: Complementarity Tests for First Releases vs. Iterative Releases

| Company Industry | Software Companies | | | | Physical Product Companies | | | |
|---|---|---|---|---|---|---|---|---|
| | (1) | (2) | (3) | (4) | (5) | (6) | (7) | (8) |
| Dependent Variables | Lead # First-Release Software Copyright | | Lead # Iterative-Release Software Copyright | | Lead # First-Release Product Trademark | | Lead # Iterative-Release Product Trademark | |
| Model | FE | FE+2SLS | FE | FE+2SLS | FE | FE+2SLS | FE | FE+2SLS |
| Prototyping | -0.442*** | -0.425*** | -0.427*** | -0.475*** | 2.256*** | 1.794*** | -0.775** | -0.774* |
| | (0.105) | (0.137) | (0.086) | (0.166) | (0.419) | (0.606) | (0.375) | (0.408) |
| Discovery-oriented AI | 0.341 | -0.329 | -0.045 | 0.689 | 1.055 | -1.659* | -0.042 | -0.348 |
| | (0.352) | (0.419) | (0.128) | (0.513) | (0.976) | (0.891) | (0.725) | (1.099) |
| Prototyping * Discovery AI | 0.437** | **1.073*** ** | 0.069 | **-0.279** | 1.433** | **7.058*** ** | 0.158 | **1.369** |
| | (0.212) | **(0.232)** | (0.056) | **(0.231)** | (0.628) | **(0.838)** | (0.685) | **(1.083)** |
| Controlled Experimentation | 0.064 | -0.112 | -0.363*** | -0.084 | -0.317 | 0.057 | -2.391*** | -2.471*** |
| | (0.116) | (0.135) | (0.097) | (0.123) | (0.547) | (0.525) | (0.514) | (0.504) |
| Optimization-oriented AI | -0.143 | 1.322 | -0.203 | -2.228 | 0.540 | 1.800 | -2.743** | -2.576** |
| | (0.521) | (1.575) | (0.362) | (2.472) | (1.499) | (2.253) | (1.065) | (1.287) |
| Controlled Experimentation * Optimization AI | 0.163 | **-0.725** | 0.535** | **2.502*** ** | 0.551 | **-2.221*** | 2.505** | **3.257*** * |
| | (0.209) | **(0.436)** | (0.267) | **(0.278)** | (1.048) | **(1.208)** | (1.147) | **(1.347)** |
| Prototyping * Optimization AI | -0.132 | 0.494 | -0.506* | -1.994*** | -0.837 | -1.533 | -1.439* | -2.491** |
| | (0.158) | (0.336) | (0.259) | (0.222) | (1.038) | (0.991) | (0.799) | (1.064) |
| Controlled Experimentation * Discovery AI | -0.468* | -1.063*** | 0.075 | 0.263 | -1.068** | -0.912 | 0.345 | -0.500 |
| | (0.263) | (0.258) | (0.092) | (0.244) | (0.423) | (1.449) | (0.276) | (1.264) |
| Controls | Firm Fixed Effect, Year Fixed Effect, Company Age, Total Number of Employees, Total Number of Job Postings, Total Number of Patents, Total Number of Certificates, Total Number of Work Copyrights, Total Number of Administrative Approvals, Total Amount of Funding | | | | | | | |
| Observations | 2,490 | 2,454 | 2,490 | 2,454 | 6,464 | 6,396 | 6,464 | 6,396 |
| R-squared | 0.502 | | 0.792 | | 0.482 | | 0.571 | |
| First-stage F-Stat | | 30.82 | | 30.82 | | 139.53 | | 139.53 |

Notes:
1. For Column (1)-(2) and (5)-(6), dependent variable is the number of first-release products in the next year. For Column (3)-(4) and (7)-(8), dependent variable is the number of iterative-release products in the next year.
2. Column (2), (4), (6) and (8) report the 2SLS results with both AI Capability and LSM instrumented.
3. Robust standard errors in parentheses: *** p<0.01, ** p<0.05, * p<0.1. Bold entries show 2SLS estimates. Statistics that are not applicable are shown in "\".

**Table 6: Discovery-oriented AI and Prototyping in Uncertain Markets**

| Company Industries | Software Companies | Physical Product Companies |
|---|---|---|
| Dependent Variable | Lead # First-Release Software Copyright | Lead # First-Release Product Trademark |
| | (1) | (2) |
| Discovery-oriented AI | 8.166*** | 12.04*** |
| | (1.244) | (2.478) |
| Prototyping | 8.989*** | 14.49*** |
| | (1.024) | (2.147) |
| Discovery-oriented AI * Prototyping | -1.083*** | -1.697** |
| | (0.137) | (0.484) |
| Market Uncertainty | 57.94*** | 28.92*** |
| | (16.11) | (7.531) |
| Market Uncertainty * Discovery AI | -11.69*** | -17.54*** |
| | (1.862) | (3.835) |
| Market Uncertainty * Prototyping | -13.84*** | -18.64*** |
| | (1.500) | (3.189) |
| Discovery AI * Prototyping * Market Uncertainty | 1.673*** | 2.302*** |
| | (0.203) | (0.761) |
| Controls | Firm Fixed Effect, Year Fixed Effect, Company Age, Total Number of Employees, Total Number of Job Postings, Total Number of Patents, Total Number of Certificates, Total Number of Work Copyrights, Total Number of Administrative Approvals, Total Amount of Funding | |
| Observations | 2,962 | 7,679 |
| R-squared | 0.470 | 0.425 |

Notes:
1. Dependent variable is the number of first-release products in the next year.
2. Discovery-oriented AI, Prototyping and Market Uncertainty are standardized.
3. Robust standard errors in parentheses: *** p<0.01, ** p<0.05, * p<0.1.

## Table 7: Different AI and LSM Effects on Product Completion Time, Business Scope, and Product Certificates

| Dependent Variable | Lead Average Completion Time (in Days) for First-Release Products | | Lead Average Completion Time (in Days) for Iterative-Release Products | | Lead # New Business Scope Entries | | Lead # Certificated Products | |
|---|---|---|---|---|---|---|---|---|
| | Software | Physical | Software | Physical | Software | Physical | Software | Physical |
| Model | (1) | (2) | (3) | (4) | (5) | (6) | (7) | (8) |
| Prototyping | 19.23*** | 13.56*** | 9.049*** | 0.333 | 0.032 | 0.036** | -0.100* | -0.007 |
| | (2.610) | (1.583) | (2.090) | (1.316) | (0.037) | (0.015) | (0.054) | (0.032) |
| Discovery AI | -2.190 | 6.206** | 3.344 | 3.803 | -0.111 | -0.074** | -0.102 | -0.127 |
| | (4.496) | (3.055) | (4.110) | (2.615) | (0.080) | (0.031) | (0.084) | (0.089) |
| Prototyping * | **-24.61*** ** | **-40.94*** ** | -1.638 | 2.497** | **0.134*** ** | **0.035** ** | 0.039 | -0.018 |
| Discovery AI | **(1.974)** | **(4.308)** | (1.608) | (1.265) | **(0.033)** | **(0.018)** | (0.026) | (0.036) |
| Controlled | 3.768 | 1.617 | 10.91*** | 9.461*** | 0.091* | 0.002 | -0.107 | -0.150*** |
| Experimentation | (2.807) | (1.741) | (2.473) | (1.530) | (0.048) | (0.016) | (0.073) | (0.039) |
| Optimization AI | 11.21 | -10.97*** | 5.773 | 4.774* | -0.201 | -0.041 | 0.277* | -0.106 |
| | (7.904) | (4.086) | (7.910) | (2.742) | (0.239) | (0.041) | (0.160) | (0.076) |
| Controlled | -13.57*** | 5.480* | **-9.239** ** | **-6.774** ** | 0.069 | 0.017 | **0.145* ** | **0.134** ** |
| Experimentation * | (2.101) | (2.970) | **(4.523)** | **(3.053)** | (0.049) | (0.011) | **(0.078)** | **(0.053)** |
| Optimization AI | | | | | | | | |
| Prototyping * | 10.15*** | 4.519 | 8.928** | 2.709 | -0.037 | -0.027** | -0.155** | -0.097** |
| Optimization AI | (1.878) | (3.092) | (4.095) | (2.085) | (0.039) | (0.013) | (0.074) | (0.046) |
| Controlled | 27.15*** | 13.25*** | -1.717 | -1.767 | -0.172*** | -0.020 | -0.005 | 0.045 |
| Experimentation * | (2.230) | (3.860) | (2.184) | (1.156) | (0.040) | (0.014) | (0.038) | (0.031) |
| Discovery AI | | | | | | | | |
| Controls | Firm Fixed Effect, Year Fixed Effect, Company Age, Total Number of Employees, Total Number of Job Postings, Total Number of Patents, Total Number of Certificates, Total Number of Work Copyrights, Total Number of Administrative Approvals, Total Amount of Funding | | | | | | | |
| Observations | 2,692 | 7,679 | 2,962 | 7,679 | 2,962 | 7,679 | 7,679 | 7,679 |
| R-squared | 0.373 | 0.423 | 0.474 | 0.550 | 0.589 | 0.702 | 0.460 | 0.667 |

Notes:

1. For Column (1)-(2) the dependent variable is the average number of days to create a first-release product (prototype) in the next year. For Column (3)-(4) the dependent variable is the average number of days to create an iterative-release product in the next year. For Column (5)-(6) the dependent variable is the number of business scope entries in the next year. For Column (7)-(8) the dependent variable is the number of certificated products in the next year.
2. Robust standard errors in parentheses: *** p<0.01, ** p<0.05, * p<0.1.

## Appendix A1: Alternative Measurement of Startup Performance

One concern about the dependent variables (software copyright and product trademark) is that they are not representative of startup's overall performances, because companies may create products that are not well received in the market. The overall performance of startups would be an ideal dependent variable but traditional metrics such as revenue and profit are not publicly available and verified by public accountants. To alleviate this concern, we replace the dependent variable using the log amount of funding in the next year in Table A1.1, as funding is one of the most important financial indicators for startups (Bardhan et al. 2013). We also use the Baidu search index as the dependent variable in Table A1.2. Baidu is the major search engine in China, and the Baidu Search Index reflects the absolute Baidu search volume of certain keywords weighted over the total search volume in the same period of time. We treat such search volume as a proxy for online traffic to the companies, and literature has shown that online traffic data can be used to predict and verify company revenue (Froot et al. 2017, Huang 2018, Zhu 2019). The results are consistent in economic and statistical significance with our earlier results that use products.

<Table A1.1 Inserted Here>

<Table A1.2 Inserted Here>

**Table A1.1: Alternative Performance Measure using Funding Amount**

| Company Industry | Software Companies | | | | | Physical Product Companies | | | | |
|---|---|---|---|---|---|---|---|---|---|---|
| | (1) | (2) | (3) | (4) | (5) | (6) | (7) | (8) | (9) | (10) |
| Dependent Variables | AI Capability | Lead Log Funding Amount | | | | AI Capability | Lead Log Funding Amount | | | |
| Model | FE | FE | FE | FE | FE+2SLS | FE | FE | FE | FE | FE+2SLS |
| AI Capability | | 0.218** | | 0.143 | 0.053 | | 0.485*** | | 0.925 | 0.265 |
| | | (0.102) | | (0.332) | (0.133) | | (0.122) | | (0.682) | (0.481) |
| LSM | 0.409*** | | 0.122 | 0.109 | 0.226 | 0.381*** | | 0.483 | 0.143 | 0.964 |
| | (0.064) | | (0.140) | (0.128) | (0.354) | (0.008) | | (0.837) | (0.369) | (1.011) |
| AI * LSM | | | | 0.901*** | **1.537*** ** | | | | 0.568*** | **1.889*** ** |
| | | | | (0.144) | **(0.431)** | | | | (0.142) | **(0.601)** |
| Controls | Company Fixed Effect, Year Fixed Effect, Company Age, Total Number of Employees, Total Number of Job Postings, Total Number of Patents, Total Number of Certificates, Total Number of Work Copyrights, Total Number of Administrative Approvals | | | | | | | | | |
| Observations | 2,962 | 2,962 | 2,962 | 2,962 | 2,962 | 7,679 | 7,679 | 7,679 | 7,679 | 7,679 |
| R-squared | 0.595 | 0.355 | 0.429 | 0.454 | \ | 0.667 | 0.316 | 0.318 | 0.591 | \ |
| $1^{st}$ stage F-stat | \ | \ | \ | \ | 62.19 | \ | \ | \ | \ | 72.07 |

Notes: 1. For Column (1) and (6), dependent variable is AI Capability in the current year. For Column (2)-(5) and (7)-(10), dependent variable is the log funding amount in the next year.
2. AI Capability is standardized, where one-standard-deviation change is associated with about 30 AI job postings, 50 AI social media articles and 15 AI news reports.
3. Column (5) and (10) report the 2SLS results with company fixed effect. 2 sets of instruments: (1) Number of AI policies issued by local governments in each year; (2) Number of news reports that a company in the same industry has adopted AI.
4. Robust standard errors in parentheses: *** p<0.01, ** p<0.05, * p<0.1.

**Table A1.2: Alternative Performance Measure using Search Index from Baidu**

| Company Industry | Software Companies | | | | | Physical Product Companies | | | | |
|---|---|---|---|---|---|---|---|---|---|---|
| | (1) | (2) | (3) | (4) | (5) | (6) | (7) | (8) | (9) | (10) |
| Dependent Variables | AI Capability | Lead Search Index | | | | AI Capability | Lead Search Index | | | |
| Model | FE | FE | FE | FE | FE+2SLS | FE | FE | FE | FE | FE+2SLS |
| AI Capability | | 0.266*** | | 0.035* | 1.312 | | 0.194* | | 0.154 | -2.513 |
| | | (0.082) | | (0.019) | (1.010) | | (0.105) | | (0.127) | (2.555) |
| LSM | 0.409*** | | 0.205 | 0.068 | 0.956 | 0.381*** | | -0.089 | 0.202 | 0.898 |
| | (0.064) | | (0.302) | (0.110) | (0.897) | (0.008) | | (0.098) | (0.288) | (1.212) |
| AI * LSM | | | | 1.523*** | **3.879*** ** | | | | 0.721*** | **2.178*** ** |
| | | | | (0.359) | **(1.222)** | | | | (0.235) | **(0.641)** |
| Controls | Company Fixed Effect, Year Fixed Effect, Company Age, Total Number of Employees, Total Number of Job Postings, Total Number of Patents, Total Number of Certificates, Total Number of Work Copyrights, Total Number of Administrative Approvals, Total Amount of Funding | | | | | | | | | |
| Observations | 2,962 | 2,962 | 2,962 | 2,962 | 2,962 | 7,679 | 7,679 | 7,679 | 7,679 | 7,679 |
| R-squared | 0.595 | 0.398 | 0.446 | 0.481 | \ | 0.667 | 0.451 | 0.502 | 0.559 | \ |
| 1st stage F-stat | \ | \ | \ | \ | 62.19 | \ | \ | \ | \ | 72.07 |

Notes: 1. For Column (1) and (6), dependent variable is AI Capability in the current year. For Column (2)-(5) and (7)-(10), dependent variable is the Baidu Search Index in the next year.
2. AI Capability is standardized, where one-standard-deviation change is associated with about 30 AI job postings, 50 AI social media articles and 15 AI news reports.
3. Column (5) and (10) report the 2SLS results with company fixed effect. 2 sets of instruments: (1) Number of AI policies issued by local governments in each year; (2) Number of news reports that a company in the same industry has adopted AI.
4. Robust standard errors in parentheses: *** p<0.01, ** p<0.05, * p<0.1.

**Appendix A2: Validation of AI and LSM Level**

There might be concerns that firms may exaggerate their AI-related activities on social media or job posting websites to attract investors. To show that the text-based measures used in the paper are also highly correlated with firms' actual AI capability, we find a subsample of companies and measure their actual AI capability by identifying their AI-related patents. If a startup is granted an AI-related patent, it shows that the company has the capability to utilize AI technologies in R&D processes (Lou and Wu 2021). So we use the cumulative number of AI-related patents to measure the firm's actual AI capability. As some startups may never apply for a patent and have 0 patent throughout the years, we only select companies with at least 1 patent to avoid potential selection biases. We classify a patent as AI-related if the patent meets any of the following two criteria: (1) The patent falls in the AI-related International Patent Classification code shown in Table A2.1 following Fujii and Managi (2018). (2) The patent contains at least one AI-related keyword in the title or abstract following Lou and Wu (2021).

<Table A2.1 Inserted Here>

Among over 85,000 patents held by our sample startups, we find about 2,000 AI-related patents. We run correlation tests in Table A2.2. The subsample contains over 4,000 observations. The dependent variable is the cumulative number of AI-related patents possessed by each company each year, and the independent variables are the text-based measurements of AI-related social media articles (WeChat), AI-related news reports and AI-related job postings. We run both simple OLS models and firm fixed-effect models. All the coefficients are statistically significant, suggesting that there is a strong correlation between the text-based measurements and the firm's actual AI capability.

<Table A2.2 Inserted Here>

**Table A2.1: Description of AI technology patent group (Fujii and Managi 2018)**

| Patent group | Description of patent group [IPC code] |
|---|---|
| Biological model | Computer systems based on biological models, including neural network models, genetic models, architectures, physical realization, learning methods, biomolecular computers, and artificial life [IPC=G06N3/00]. |
| Knowledge-based model | Computer systems that utilize knowledge-based models, including knowledge engineering, knowledge acquisition, extracting rules from data, and inference methods or devices [IPC=G06N5/00]. |
| Specific mathematical model | Computer systems based on specific mathematical models, including fuzzy logic, physical realization, chaos models or non-linear system models, and probabilistic networks [G06/N7/00]. |
| Other AI technology | Subject matter not provided for previously described groups of the G06N3 subclass, including quantum computers, learning machines, and molecular computers [G06/N99/00]. |

**Table A2.2: Correlation between Text-Measured AI Capabilities and AI-related Patents**

| DV: Cumulative Number of AI-related Patent | (1) | (2) | (3) | (4) | (5) | (6) |
|---|---|---|---|---|---|---|
| Model | OLS | Firm FE | OLS | Firm FE | OLS | Firm FE |
| AI-related WeChat | 0.0243*** | 0.0369*** | | | | |
| | (0.0062) | (0.00647) | | | | |
| AI-related News | | | 0.175*** | 0.182*** | | |
| | | | (0.0501) | (0.0344) | | |
| AI-related Jobs | | | | | 0.0512*** | 0.0788*** |
| | | | | | (0.0171) | (0.0133) |
| Observations | 4,413 | 4,413 | 4,413 | 4,413 | 4,413 | 4,413 |
| R-squared | 0.006 | 0.687 | 0.025 | 0.696 | 0.062 | 0.686 |

Robust standard errors in parentheses
*** p<0.01, ** p<0.05, * p<0.1

As for the measurement of LSM, we adopt two ways to verify our empirical measurement of LSM. First, we measure whether the startup has received funding from government or state-owned funds. Prior literature has documented that startups receiving support from public institutions are more likely to produce and follow business plans (Honig and Karlsson 2004), because government funding usually requires a rigorous business plan and subsequent execution from the invested startup. By contrast, startups funded by venture capital firms are more flexible to pivot and use the LSM. In Figure A2.1, we show the distribution of the LSM level of startups that have or have not received funding from state-owned funds. Consistent in theory, companies without state-owned funding have a significantly higher level of LSM in comparison to companies with state-owned funding.

<Figure A2.1 Inserted Here>

We also verify our LSM measure using the concentration level of investor voting rights. Prior literature has documented that it is usually harder for startups to pivot when the voting rights of the investors are dispersed among shareholders because it is harder to reach a consensus at the top management team (Brinckmann et al. 2010). We calculate the concentration of voting rights each year for a firm using the Herfindahl-Hirschman Index (HHI) based on the proportion of shares each investor has in the firm. A higher HHI suggests that a few people hold the majority of voting rights in the firm. In Figure A2.2, we show the distribution of the LSM level of startups that have an above-median HHI versus the startups that have a below-median HHI. Companies with more concentrated voting rights have a significantly higher level of LSM than firms with more dispersed voting rights. These results help validate our empirical measurement of LSM level.

<Figure A2.2 Inserted Here>

**Figure A2.1: LSM Level for Startups with or without Government Funding**

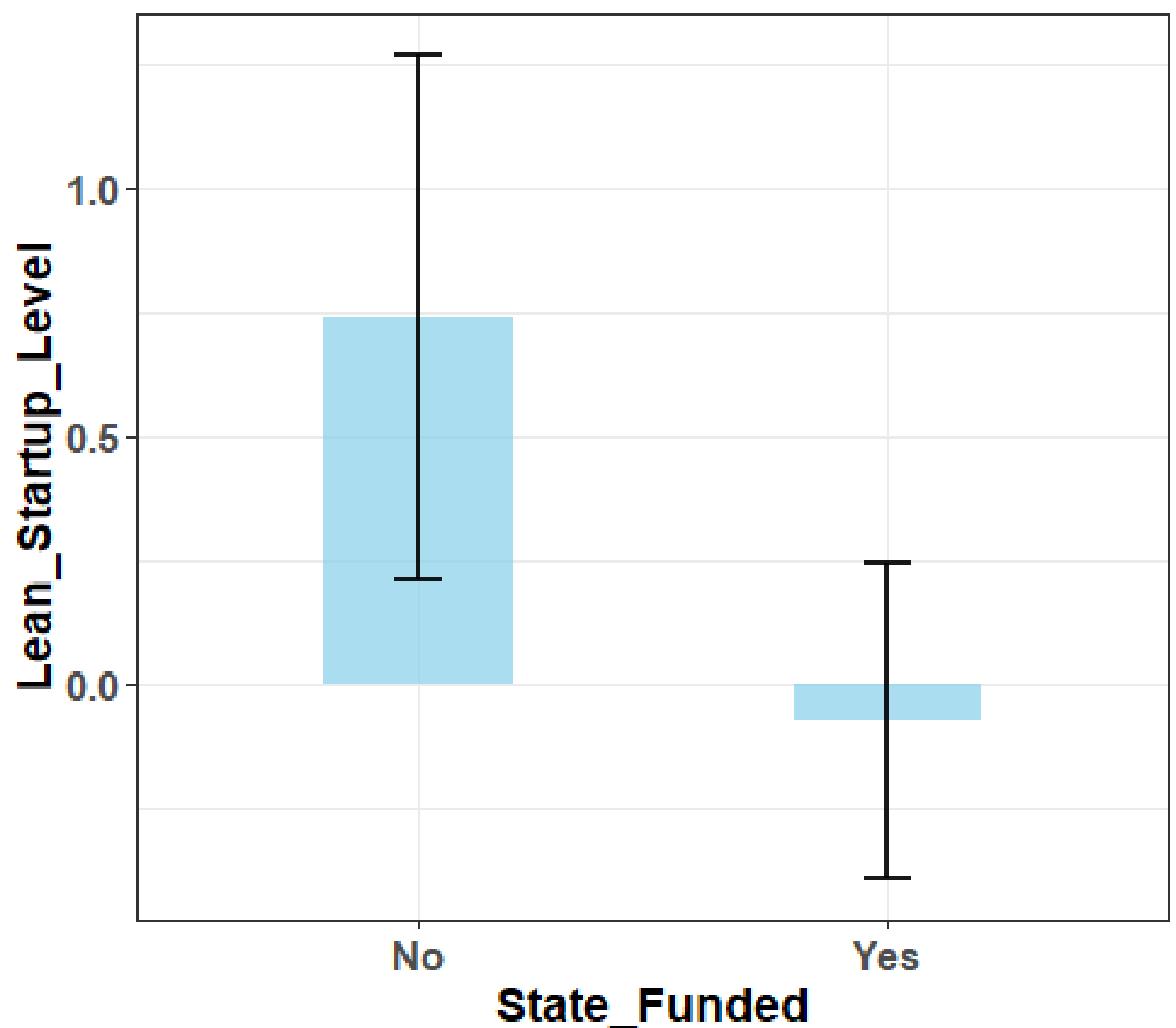

**Figure A2.2: LSM Level for Startups above or below the Median of Share HHI**

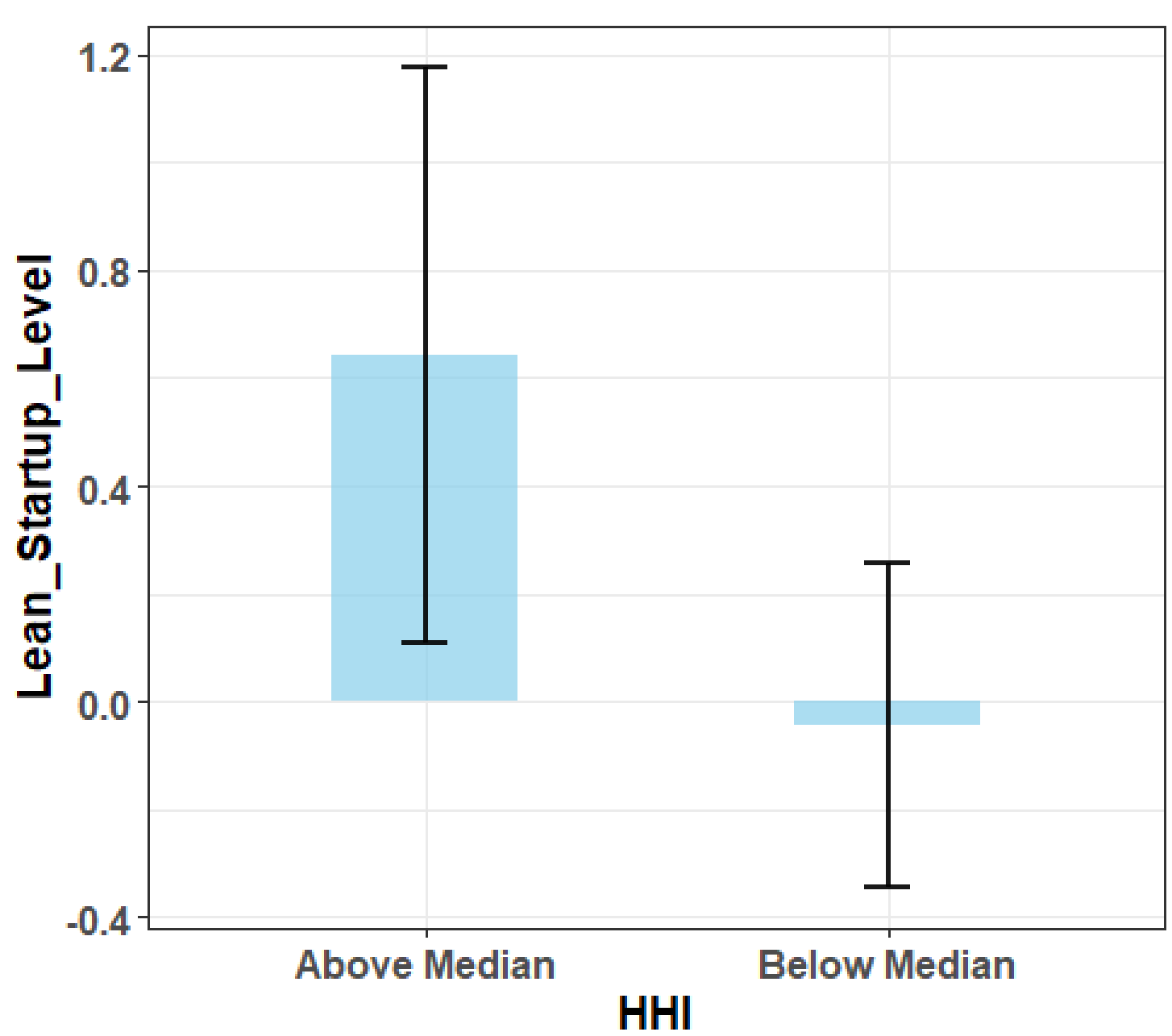

## Appendix A3: Table of Correlations

### Table A3.1: Correlations between the Main Variables

| | (1) | (2) | (3) | (4) | (5) | (6) | (7) | (8) | (9) |
|---|---|---|---|---|---|---|---|---|---|
| (1) Lead Number of Software Copyrights | 1.00 | | | | | | | | |
| (2) Lead Number of Trademarks | 0.16 | 1.00 | | | | | | | |
| (3) Lead Number of First-Release Software Copyrights | 0.64 | 0.14 | 1.00 | | | | | | |
| (4) Lead Number of First-Release Trademarks | 0.12 | 0.68 | 0.11 | 1.00 | | | | | |
| (5) Lead Number of Iterative-Release Software Copyrights | 0.48 | 0.20 | 0.38 | 0.16 | 1.00 | | | | |
| (6) Lead Number of Iterative-Release Trademarks | 0.17 | 0.67 | 0.14 | 0.63 | 0.21 | 1.00 | | | |
| (7) Lead Number of Certificates | 0.12 | 0.13 | 0.10 | 0.09 | 0.12 | 0.13 | 1.00 | | |
| (8) AI Capability | 0.22 | 0.17 | 0.19 | 0.17 | 0.32 | 0.16 | 0.08 | 1.00 | |
| (9) Discovery-oriented AI | 0.20 | 0.12 | 0.41 | 0.20 | 0.37 | 0.16 | 0.39 | 0.31 | 1.00 |
| (10) Optimization-oriented AI | 0.10 | 0.37 | 0.06 | 0.09 | 0.05 | 0.06 | 0.14 | 0.37 | 0.45 |
| (11) LSM Level | 0.20 | 0.41 | 0.18 | 0.34 | 0.29 | 0.41 | 0.08 | 0.39 | 0.18 |
| (12) Prototyping | 0.08 | 0.15 | 0.06 | 0.10 | 0.13 | 0.17 | 0.01 | 0.12 | 0.06 |
| (13) Controlled Experimentation | 0.17 | 0.14 | 0.14 | 0.10 | 0.30 | 0.15 | 0.08 | 0.19 | 0.14 |
| (14) Age | 0.12 | 0.06 | 0.09 | 0.02 | 0.35 | 0.07 | 0.11 | 0.14 | 0.09 |
| (15) Number of Branches | 0.17 | 0.22 | 0.14 | 0.16 | 0.24 | 0.23 | 0.07 | 0.08 | 0.14 |
| (16) Number of Subsidiaries | 0.03 | 0.07 | 0.02 | 0.04 | 0.03 | 0.08 | 0.01 | 0.01 | 0.02 |
| (17) Total Number of Job Posts | 0.16 | 0.23 | 0.14 | 0.18 | 0.27 | 0.24 | 0.06 | 0.41 | 0.14 |
| (18) Total Number of Employees | 0.20 | 0.33 | 0.17 | 0.28 | 0.28 | 0.32 | 0.08 | 0.16 | 0.17 |
| (19) Market Uncertainty | -0.16 | -0.20 | -0.14 | -0.14 | -0.27 | -0.21 | -0.07 | -0.09 | -0.14 |
| (20) Market Competition | 0.13 | 0.08 | 0.12 | 0.06 | 0.24 | 0.09 | 0.09 | 0.12 | 0.12 |
| (21) Total Number of Administrative Approvals | 0.13 | 0.58 | 0.10 | 0.49 | 0.34 | 0.58 | 0.11 | 0.30 | 0.10 |
| (22) Total Number of Patents | 0.13 | 0.14 | 0.11 | 0.12 | 0.28 | 0.13 | 0.11 | 0.28 | 0.11 |
| (23) Total Number of Certificates | 0.06 | 0.09 | 0.05 | 0.06 | 0.14 | 0.10 | 0.47 | 0.07 | 0.05 |
| (24) Total Number of Work Copyrights | 0.01 | 0.04 | 0.00 | 0.02 | 0.06 | 0.05 | 0.01 | 0.02 | 0.00 |
| (25) Total Funding Rounds | 0.08 | 0.15 | 0.06 | 0.10 | 0.13 | 0.17 | 0.01 | 0.12 | 0.06 |
| (26) Total Funding Amount | 0.17 | 0.14 | 0.14 | 0.10 | 0.30 | 0.15 | 0.08 | 0.19 | 0.14 |
| | | | | | | | | | |
| **Continue from above** | (10) | (11) | (12) | (13) | (14) | (15) | (16) | (17) | (18) |
| (10) Internally Focused AI | 1.00 | | | | | | | | |
| (11) LSM Level | 0.39 | 1.00 | | | | | | | |
| (12) Prototyping | 0.08 | 0.05 | 1.00 | | | | | | |
| (13) Controlled Experimentation | 0.18 | 0.45 | 0.43 | 1.00 | | | | | |
| (14) Age | 0.09 | 0.50 | 0.13 | 0.09 | 1.00 | | | | |
| (15) Number of Branches | 0.26 | 0.22 | 0.23 | 0.14 | 0.35 | 1.00 | | | |
| (16) Number of Subsidiaries | 0.38 | 0.15 | 0.11 | 0.06 | 0.16 | 0.19 | 1.00 | | |
| (17) Total Number of Job Posts | 0.20 | 0.24 | 0.09 | 0.14 | 0.16 | 0.15 | 0.17 | 1.00 | |
| (18) Total Number of Employees | 0.15 | 0.24 | 0.16 | 0.01 | 0.18 | 0.29 | 0.09 | 0.18 | 1.00 |
| (19) Market Uncertainty | -0.27 | -0.20 | -0.26 | -0.07 | -0.51 | -0.36 | 0.25 | -0.15 | -0.02 |
| (20) Market Competition | 0.56 | 0.42 | 0.06 | 0.01 | 0.07 | 0.08 | 0.05 | 0.11 | 0.15 |
| (21) Total Number of Administrative Approvals | 0.08 | 0.15 | 0.07 | 0.00 | 0.01 | 0.04 | 0.02 | 0.01 | 0.04 |
| (22) Total Number of Patents | 0.04 | 0.04 | 0.24 | 0.03 | 0.26 | 0.27 | 0.27 | 0.24 | 0.35 |
| (23) Total Number of Certificates | 0.28 | 0.35 | 0.19 | 0.08 | 0.18 | 0.22 | 0.08 | 0.09 | 0.12 |
| (24) Total Number of Work Copyrights | 0.21 | 0.12 | 0.16 | 0.10 | 0.25 | 0.19 | 0.17 | 0.18 | 0.47 |
| (25) Total Funding Rounds | 0.27 | 0.47 | 0.06 | 0.10 | 0.13 | 0.17 | 0.01 | 0.12 | 0.15 |
| (26) Total Funding Amount | 0.08 | 0.15 | | | | | | | |
| | | | | | | | | | |
| **Continue from above** | (19) | (20) | (21) | (22) | (23) | (24) | (25) | (26) | |
| (19) Market Uncertainty | 1.00 | | | | | | | | |
| (20) Market Competition | -0.47 | 1.00 | | | | | | | |
| (21) Total Number of Administrative Approvals | -0.15 | 0.14 | 1.00 | | | | | | |
| (22) Total Number of Patents | -0.10 | 0.02 | 0.01 | 1.00 | | | | | |
| (23) Total Number of Certificates | -0.34 | 0.28 | 0.13 | 0.07 | 1.00 | | | | |
| (24) Total Number of Work Copyrights | -0.22 | 0.20 | 0.02 | 0.00 | 0.12 | 1.00 | | | |
| (25) Total Funding Rounds | -0.25 | 0.17 | 0.08 | 0.06 | 0.29 | 0.20 | 1.00 | | |
| (26) Total Funding Amount | -0.01 | 0.04 | 0.00 | 0.02 | 0.06 | 0.05 | 0.26 | 1.00 | |

## Appendix A4: Time Variation of AI-related Policies

The number of AI-related policies as an instrumental variable varies on region-year level. As for the concerns about limited variations in AI policies in some regions over time, all the first-stage instrumental variable regression results have F-statistics greater than 20, suggesting that there is sufficient variation in the instrumental variable to identify the independent variable. For example, in Table A4.1 we show the first-stage instrumental variable regression result using AI capability as the dependent variable and AI-related policies as the independent variable.

**Table A4.1: First-Stage Instrumental Variable Result**

| Company Industries | Software Companies | Physical Product Companies |
|---|---|---|
| Dependent Variable: AI Capability | (1) | (2) |
| Provincial AI Policies | 0.0022*** | 0.0038*** |
| | (0.0004) | (0.0002) |
| Other Controls | Company Fixed Effect, Year Fixed Effect, Company Age, Total Number of Employees, Total Number of Job Postings, Total Number of Patents, Total Number of Certificates, Total Number of Work Copyrights, Total Number of Administrative Approvals, Total Amount of Funding | |
| Observations | 2,962 | 7,679 |
| F-Stat | 325.19 | 1607.2 |

Notes:
1. Dependent variable is AI Capability in each startup each year.
2. Independent variable is the number of AI-related policies published by local Chinese governments in each year.
3. Robust standard errors in parentheses: *** p<0.01, ** p<0.05, * p<0.1.

Moreover, we also conduct one-way ANOVA to calculate the between-region variance and within-region variance for AI-related policies in 26 provincial-level regions across 10 years in our sample. The within-region variance shows the variation of AI-related policies within the same region over time. The results are in Table A4.2:

**Table A4.2: One-Way ANOVA for AI-related Policies**

| **Source of Variation** | **Sum of Squares** | **Degree of Freedom** | **Mean Square** | **F-statistics** | **Prob>F** |
|---|---|---|---|---|---|
| Between Regions | 5,330.6 | 25 | 213.2 | 7.56 | 0.00 |
| Within Regions | 6,607.9 | 234 | 28.2 | | |
| | | | | | |
| Total | 11,938.5 | 259 | | | |

The results show that between-region variance explains 45% (5330.6/11938.5) of the total variance of AI-related policies, while within-region variance explains 55%. Such results suggest that the variation of AI-related policies over time is high in general. We have also revised the manuscript to provide more details about the variation of the instrumental variables.

**Appendix A5: Reduced-Form Instrumental Variable Test**

A reasonable concern about the instrumental variable is that the AI policies published by local governments may be correlated with regional socio-economic levels and thereby correlated with unobserved firm qualities. The number of AI-related news about companies in the same industry may also be correlated with unobserved market-level dynamics and firm qualities. If that is the case, then we should also observe that instrumental variables are associated with startup innovation activities even when the focal startup has not adopted AI capability. We show the reduced-form instrumental variable test in Table A5.1 following the procedures in Martin and Yurukoglu (2017) and Wang et al. (2023). Specifically, we create a binary variable of AI Adoption by identifying the earliest year when the company has any AI-related patent, job posting, social media article, or news report. We then separate our sample into AI adopters versus non-adopters and use the instrumental variables as the independent variable directly in the regression. The results show that the instruments are only significantly associated with startup innovation activities when startups have adopted AI, suggesting that our instrumental variable only correlates with the dependent variable through AI adoption.

<Table A5.1 Inserted Here>

**Table A5.1: Reduced-Form Instrumental Variable Test**

| AI Adopters or Not | AI Adopters | | Non-Adopters | |
|---|---|---|---|---|
| Company Industries | Software Companies | Physical Product Companies | Software Companies | Physical Product Companies |
| Dependent Variable | Lead # Software Copyright | Lead # Product Trademark | Lead # Software Copyright | Lead # Product Trademark |
| | (1) | (2) | (3) | (4) |
| Provincial AI Policies | 0.025*** | 0.076*** | -0.003 | 0.015 |
| | (0.005) | (0.026) | (0.007) | (0.011) |
| AI News from Companies in the Same Industry | 0.011** | 0.031* | 0.015 | 0.006 |
| | (0.006) | (0.017) | (0.010) | (0.006) |
| Controls | Company Fixed Effect, Year Fixed Effect, Company Age, Total Number of Employees, Total Number of Job Postings, Total Number of Patents, Total Number of Certificates, Total Number of Work Copyrights, Total Number of Administrative Approvals, Total Amount of Funding | | | |
| Observations | 1,262 | 2,014 | 1,228 | 4,450 |
| R-squared | 0.189 | 0.254 | 0.178 | 0.254 |

Notes: 1. Dependent variable is the number of new products in the next year.
2. Independent variable is the number of AI-related policies published by local Chinese governments in each year.
3. Robust standard errors in parentheses: *** p<0.01, ** p<0.05, * p<0.1.

## Appendix A6: Skewness of Dependent Variable and Independent Variables

In the sample, some variables, such as the number of software copyrights or the number of employees, are skewed because startups usually grow very fast. To alleviate the concerns about the skewness of data we have run additional robustness tests, including using the logged value of the dependent variable and also using coarsened exact matching:

(1) To alleviate the concern about the skewness of the dependent variable (number of product innovations), we tested the regression results after taking logs for the dependent variables. We show the main regression results in Table A6.1 and they are consistent with the results using unlogged values.

(2) To alleviate the concern about the skewness of the independent variables such as the job posts and the number of employees, we also run coarsened exact matching on the sample to obtain a more balanced dataset (Blackwell et al. 2009). For software startups, we divide our sample into two subsamples, with the cutting point being the median of AI adoption level. We then use coarsened exact matching to find observations that are more balanced between the two subsamples. About 200 observations are dropped because of extreme values. We then run the main regression results only using the matched sample. The same procedures are also applied to physical product startups. The results are shown in Table A6.2 and are consistent with the results using the original sample.

**Table A6.1: Robustness Test using Logged Dependent Variable**

| Company Industry | Software Companies | | | | | Physical Product Companies | | | | |
|---|---|---|---|---|---|---|---|---|---|---|
| | (1) | (2) | (3) | (4) | (5) | (6) | (7) | (8) | (9) | (10) |
| Dependent Variables | AI Capability | Lead Log # Software Copyright | | | | AI Capability | Lead Log # Product Trademark | | | |
| Model | FE | FE | FE | FE | FE+2SLS | FE | FE | FE | FE | FE+2SLS |
| AI Capability | | 0.092** | | 0.056 | -0.120*** | | 0.033*** | | -0.256*** | -0.454*** |
| | | (0.041) | | (0.051) | (0.020) | | (0.010) | | (0.066) | (0.106) |
| LSM | 0.409*** | | -0.010 | 0.004 | 0.568*** | 0.381*** | | 0.185*** | 0.701 | 0.148*** |
| | (0.064) | | (0.024) | (0.022) | (0.117) | (0.008) | | (0.017) | (0.921) | (0.025) |
| AI * LSM | | | | 0.105*** | **0.242**** | | | | 0.353*** | **0.661***** |
| | | | | (0.027) | **(0.119)** | | | | (0.099) | **(0.099)** |
| Controls | Company Fixed Effect, Year Fixed Effect, Company Age, Total Number of Employees, Total Number of Job Postings, Total Number of Patents, Total Number of Certificates, Total Number of Work Copyrights, Total Number of Administrative Approvals, Total Amount of Funding | | | | | | | | | |
| Observations | 2,962 | 2,962 | 2,962 | 2,962 | 2,962 | 7,679 | 7,679 | 7,679 | 7,679 | 7,679 |
| R-squared | 0.595 | 0.341 | 0.339 | 0.344 | \ | 0.667 | 0.277 | 0.290 | 0.313 | \ |
| 1st stage F-stat | \ | \ | \ | \ | 62.19 | \ | \ | \ | \ | 72.07 |

Notes: 1. For Column (1) and (6), dependent variable is AI Capability in the current year. For Column (2)-(5) and (7)-(10), dependent variable is the logged number of new products (total number + 1) in the next year.
2. AI Capability is standardized, where one-standard-deviation change is associated with about 30 AI job postings, 50 AI social media articles and 15 AI news reports.
3. Column (5) and (10) report the 2SLS results with both AI Capability and LSM instrumented.
4. Robust standard errors in parentheses: *** p<0.01, ** p<0.05, * p<0.1.

## Table A6.2: Robustness Test using Coarsened Exact Matching Sample

| Company Industry | Software Companies | | | | | Physical Product Companies | | | | |
|---|---|---|---|---|---|---|---|---|---|---|
| | (1) | (2) | (3) | (4) | (5) | (6) | (7) | (8) | (9) | (10) |
| Dependent Variables | AI Capability | Lead Log # Software Copyright | | | | AI Capability | Lead Log # Product Trademark | | | |
| Model | FE | FE | FE | FE | FE+2SLS | FE | FE | FE | FE | FE+2SLS |
| AI Capability | | 0.422*** | | -0.550 | -5.726*** | | 3.506** | | -5.398*** | -32.94*** |
| | | (0.079) | | (0.431) | (1.272) | | (1.646) | | (1.786) | (6.169) |
| LSM | 0.404*** | | -0.227 | -0.523*** | -1.344*** | 0.391*** | | 3.881*** | 2.569*** | -1.472 |
| | (0.063) | | (0.174) | (0.189) | (0.274) | (0.010) | | (0.808) | (0.817) | (1.252) |
| AI * LSM | | | | 0.180*** | **0.985*** ** | | | | 4.089*** | **18.68*** ** |
| | | | | (0.0476) | **(0.191)** | | | | (0.365) | **(3.119)** |
| Controls | Company Fixed Effect, Year Fixed Effect, Company Age, Total Number of Employees, Total Number of Job Postings, Total Number of Patents, Total Number of Certificates, Total Number of Work Copyrights, Total Number of Administrative Approvals, Total Amount of Funding | | | | | | | | | |
| Observations | 2,759 | 2,759 | 2,759 | 2,759 | 2,759 | 7,492 | 7,492 | 7,492 | 7,492 | 7,492 |
| R-squared | 0.598 | 0.380 | 0.380 | 0.385 | \ | 0.678 | 0.419 | 0.421 | 0.433 | \ |
| 1st stage F-stat | \ | \ | \ | \ | 82.44 | \ | \ | \ | \ | 66.18 |

Notes: 1. For Column (1) and (6), dependent variable is AI Capability in the current year. For Column (2)-(5) and (7)-(10), dependent variable is the number of new products in the next year.

2. AI Capability is standardized, where one-standard-deviation change is associated with about 30 AI job postings, 50 AI social media articles and 15 AI news reports.

3. Column (5) and (10) report the 2SLS results with both AI Capability and LSM instrumented.

4. Robust standard errors in parentheses: *** p<0.01, ** p<0.05, * p<0.1.